\documentclass[preprint,12pt]{elsarticle}

\usepackage{amssymb}
\usepackage[table]{xcolor}
\usepackage[T1]{fontenc}
\usepackage[utf8]{inputenc}
\usepackage{multirow}
\usepackage{array}
\usepackage{longtable}
\usepackage{booktabs}
\usepackage{hyperref}
\usepackage{subcaption}
\usepackage{tikz}
\usepackage[most]{tcolorbox}

\newcommand{\rotl}[1]{\rotatebox{90}{\textbf{\footnotesize #1}}}
\newcommand{\Y}{\cellcolor{green!25}\checkmark}
\newcommand{\Pa}{\cellcolor{yellow!25}$\pm$}
\newcommand{\N}{\cellcolor{red!25}}

\journal{Data \& Knowledge Engineering}

\begin{document}

\begin{frontmatter}



\title{Robotic Process Automation Using Process Mining -- A Systematic Literature Review\tnoteref{t1}}
\tnotetext[t1]{This work was supported by the Discovery program of the Natural Sciences and Engineering Research of Canada, and by the University of Ottawa, Canada.}


\author[1]{Najah Mary El-Gharib}
\ead{nelgh031@uottawa.ca}
\author[1]{Daniel Amyot\corref{cor1}}
\ead{damyot@uottawa.ca}
\cortext[cor1]{Corresponding author}

\affiliation[1]{
    organization={School of Electrical Engineering and Computer Science, University of Ottawa},
    addressline={800 King Edward St}, 
    city={Ottawa},
    postcode={K1N 6N5}, 
    state={ON},
    country={Canada}
}

\begin{abstract}
Process mining (PM) aims to construct, from event logs, process maps that can help discover, automate, improve, and monitor organizational processes. Robotic process automation (RPA) uses software robots to perform some tasks usually executed by humans. It is usually difficult to determine what processes and steps to automate, especially with RPA. PM is seen as one way to address such difficulty. This paper aims to assess the applicability of process mining in accelerating and improving the implementation of RPA, along with the challenges encountered throughout project lifecycle.

A systematic literature review was conducted to examine the approaches where PM techniques were used to understand the as-is processes that can be automated with software robots. Seven databases were used to identify papers on this topic. A total of 32 papers, all published since 2018, were selected from 605 unique candidate papers and then analyzed. 

There is a steady increase in the number of publications in this domain, especially during the year 2022, which suggests a raising interest in the combined use of PM with RPA. The literature mainly focuses on the methods to record the events that occur at the level of user interactions with the application, and on the preprocessing methods that are needed to discover routines with the steps that can be automated. Important challenges are faced with preprocessing such event logs, and many lifecycle steps of automation projects are weakly supported by existing approaches suggesting corresponding research areas in need of further attention.
\end{abstract}


\begin{highlights}
\item Combining process mining (PM) and RPA offers unique process management opportunities.
\item PM techniques must better support discovering processes based on UI logs for RPA.
\item Tools need to better integrate PM and RPA features together, in a synergetic way.
\item Challenges common to PM and RPA remain, e.g., with data gathering and preprocessing. 
\end{highlights}


\end{frontmatter}

\section{Introduction}
\label{sect:introduction}
In the context of digital transformation, many organizations are automating their manual processes to improve performance, save costs, and minimize errors while executing these processes. However, there is currently a large amount of guess work in assessing the processes that can be automated and in monitoring their actual improvement. In the past five years, there has been a steep increase in the use of \emph{robotic process automation} (RPA) in organizations, together with related tool support~\cite{Aalst2018}. RPA, which uses software robots to automate human tasks, has often been applied in the areas of administration and finance. RPA is frequently implemented in cases where high volumes of data are processed through repetitive tasks that can be automated. As reported by \citet{RPAtools2023}, the market of RPA solutions includes over 53 vendors who develop RPA tools that provide different functionalities to automate office tasks in an intelligent way. Understanding processes is key to automating them, but organizations often lack a deep understanding of their as-is processes and the way they are being executed in reality.

\emph{Process mining} (PM) is an emerging technology aiming to generate process maps (also known as process models) from event logs and discover valuable insights towards process understanding and improvement. PM takes event logs collected from information systems as input and produces process maps using discovery algorithms.

The implementation of a successful and reliable automated process requires understanding the detailed activities that compose that process. In order to configure RPA technologies, define the process paths and steps, and specify what conditions trigger certain actions to occur, the process needs to be modeled with a sufficient level of detail. 
Implementing RPA can be costly and time consuming, especially when the selected process is too complex with high variations. Identify the suitable processes for RPA implementations is critical step.
Often, discovering as-is processes is done manually through a mix of interviews, workshops, documentation analysis, and desktop monitoring. This approach enables engineers to define the activities that a software robot has to perform, but it is time consuming and multiple errors can occur in understanding and automating the process steps. An approach is needed that enables capturing the detailed and precise information about a given process, including its possible variations. Given its nature, process mining can likely help in that context. \citet{Aalst2018} were among the first researchers to observe a link between process mining and robotic process automation. \citet{Beetz2019RoboticPA}, as part of their work on an evaluation model to select processes that are suitable for automation, highlighted that special attention should be given to process mining techniques when evaluating RPA implementations. In addition, \citet{RPASLR2020} observed that less than 15\% of 14 evaluated commercial RPA tools partially support the kind of features needed for analysing which processes to automate.

There is an increasingly important need for practitioners and researchers who want to understand how best to combine RPA with PM for solving real process automation issues to understand the current status of how RPA and PM can effectively be combined. Industrial surveys and white papers on that topic start to emerge, such as Deloitte's global process mining recent survey~\cite{Deloitte2023}, which highlights that 70\% of participants agreed that they are using process mining in conjunction with RPA applications. However, such reports abstract away the academic perspective, for which there is now a critical amount of published evidence and lessons learned related to techniques, tools, and challenges related to how PM can support RPA. From an academic perspective, there has been a significant increase in the number of publications on the interested for PM and RPA over the last 2 years.

The objective of this systematic literature review is to determine how process mining can complement robotic process automation to accelerate and improve RPA implementation. More specifically, this review aims to answer the following research questions:

\begin{itemize}
\item
  \textbf{RQ1:} How are process mining techniques applied to accelerate
  and improve robotic process automation implementation?
\item
  \textbf{RQ2:} Which tools are used to apply both process mining and
  robotic process automation in an integrated way?
\item
  \textbf{RQ3:} What are the challenges encountered when combining
  process mining with robotic process automation?
\end{itemize}

Out of the 605 papers that were returned by seven search engines, 45 papers were assessed for eligibility and 32 relevant papers were selected, reviewed, and analyzed to answer the research questions described above. Although there exist several systematic literature reviews on the topic of RPA~\cite{Beetz2019RoboticPA,RPASLR2019,RPASLR2020,CostaRPASLR2022,BPMRPA2021,Syed2020RoboticPA,Wewerka2021}, none of them covers the intersection of process mining and RPA, which is the focus of our literature review. These systematic literature reviews focused on reporting the state-of-the-art of RPA, including themes on RPA meanings, differences related to technologies, criteria of selecting process tasks, RPA use cases, RPA effects, methods for RPA projects, and challenges for future research \cite{Wewerka2021, Syed2020RoboticPA, RPASLR2020, RPASLR2019}.
\citet{RPASLR2020} also assessed the industrial literature and highlighted the absence of technological solutions and tools to discover the best candidate processes of an organization to be automated. Additionally, \citet{Beetz2019RoboticPA} developed a process evaluation model to identify RPA suitable processes as a result of the systematic literature review and interviews with experts.
Even though this review highlighted that at the pre-selection phase of an RPA project process mining can be used as a support tool, this SLR did not focus on how process mining techniques specifically can be utilized. Only one SLR aims to assess the difference between RPA and Business Process Management Systems (BPMSs) and how RPA is used in business practice \cite{RPASLR2019}.

This review will benefit researchers and practitioners interested in obtaining an overview of existing research and applications on how to improve robotic process automation projects by combining them with process mining discovery approaches. Additionally, this review discusses how to improve and tailor existing process mining techniques to be applied in the RPA domain.

The organization of this paper is as follows. Section~\ref{sect:overview} presents an overview of process mining and robotic process automation, together with an example. Section~\ref{sect:methodology} introduces the literature review methodology and the selected papers. Section~\ref{sect:results} presents the results of the analysis of the 32 selected papers, and Section~\ref{sect:answers} answers the three research questions. Threats to validity are discussed in Section~\ref{sect:threats}, while Section~\ref{sect:conclusion} concludes.

\section{Overview of Process Mining and Robotic Process Automation}
\label{sect:overview}
This section provides an overview of the two technologies of interest here (PM and RPA), together with an illustrative example.

\subsection{Overview of Process Mining}
\label{sect:overviewPM}
\begin{figure}
    \centering
    \includegraphics[width=\textwidth]{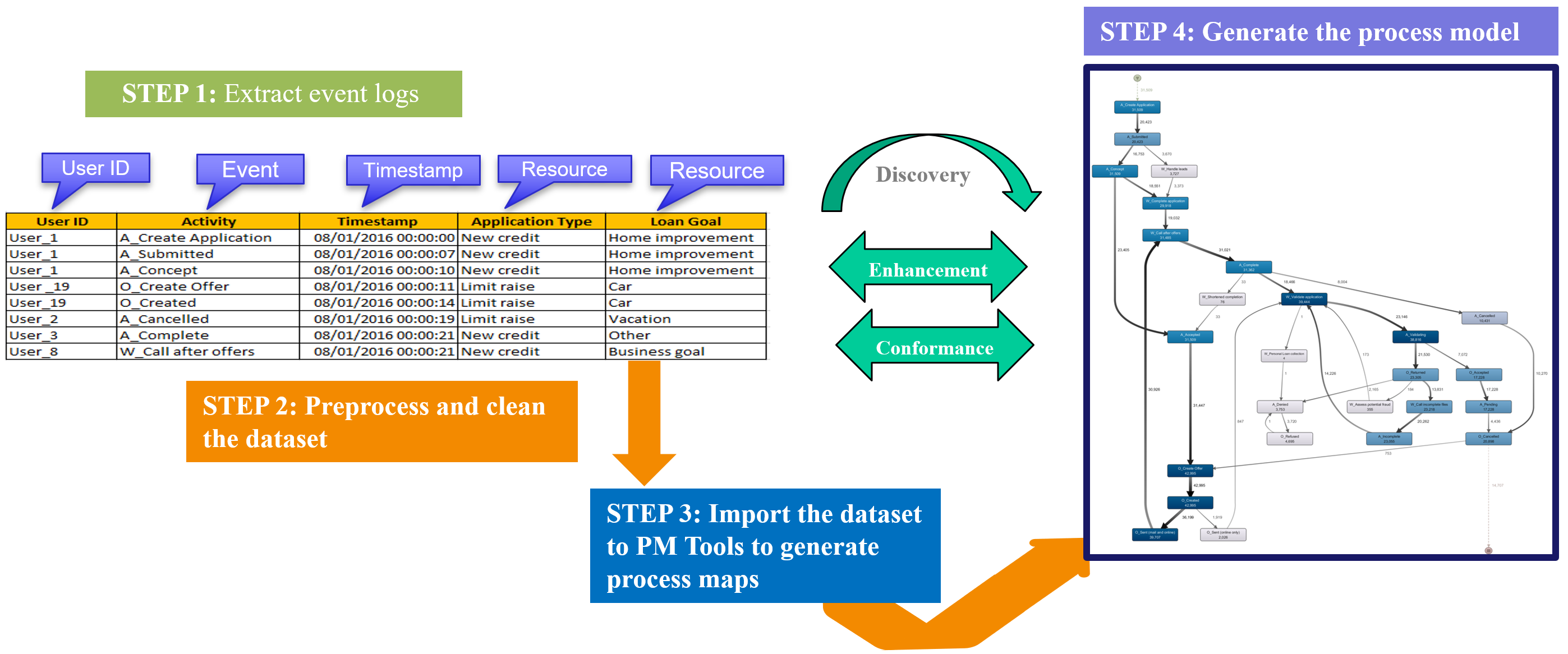}
    \caption{Overview of Process Mining}
    \label{fig:overviewPM}
\end{figure}

Process mining is an emerging analytical discipline providing novel techniques to discover, monitor, and improve processes by extracting valuable knowledge and information from event logs available in information systems. Process mining offers evidence-based insights that are derived from actual data, in order to help organizations audit, analyze, and improve their existing business processes by answering both compliance-related and performance-related questions. Figure~\ref{fig:overviewPM} shows the four main steps commonly followed to apply process mining, starting with extracting event logs from information systems~(1), preprocessing and cleaning the dataset~(2), and importing the event logs to process mining tools~(3), which are finally used to generate process maps/models~(4).

Event logs extracted from information systems are the starting point of PM~\cite{PMbook}. The main assumption about the event log structure is that it contains data related to a single process. Each event in a log refers to an activity, which is a well-defined step in a process, and is related to a particular case identifying a process instance. The events belonging to a case are usually sorted and contain a timestamp enabling sorting. A case is a specific instance of a process and a process contains several cases. Each case has a unique identifier referred as case ID, which determines the scope of the process instance, i.e., where the instance starts and where it ends.

In general, process mining activities are categorized into three main types: discovery, conformance, and enhancement~\cite{PMbook}. \emph{Process discovery} takes an event log as input and produces a process model as output, without a priori information. It is the most popular process mining activity, and also the focus of this paper. 

%\emph{Conformance checking} compares an existing process model with an event log generated by an executing process. This activity is used to check whether the reality, as recorded in the log, conforms to the desired or required model~\cite{PMdrift2022}. \emph{Enhancement} improves an existing process model by using information about the actual process recorded in the event logs previously discovered using PM techniques~\cite{PMbook}.

There are several process mining algorithms that have been proposed by researchers in this domain, as reviewed by \citet{GARCIA2019}. These algorithms target different output formalisms (precedence graphs, Petri Nets, BPMN models, and others), various types of annotations on nodes and edges (e.g., frequencies and costs), whether concurrency is mined or not, and other such distinctive characteristics. These algorithms and their benefits are however outside the scope of our review.

%Some of the most popular discovery algorithms include the $\alpha$-algorithm and Heuristic Mining~\cite{PMbook}). The \emph{$\alpha$-algorithm} takes an event log as an input and generates a Petri net model, including concurrency relationships, displaying the behavior included in the input event logs. The \emph{Heuristic Mining} algorithm produces a model without concurrency but takes the frequencies of events and sequences into account when generating the process model. 

There are several tools that can accept event logs as inputs and generate process maps using process mining algorithms. The ProM framework~\cite{ProM} is an open-source tool implemented in Java, supported by a leading academic group from the Eindhoven Institute of Technology. It supports multiple techniques for process discovery, conformance checking, organizational mining, social network analysis, and decision mining. PM4PY~\cite{PM4PY} is another open-source tool, this time written in Python. According to Gartner~\cite{Gartner-PM-2020}, there are several other commercial tools in the market including Apromore~\cite{Apromore}, Celonis~\cite{Celonis}, Disco~\cite{Disco}, Microsoft Minit~\cite{Minit} (now part of Microsoft Power Automate), Signavio Process Intelligence~\cite{Signavio}, and UiPath Process Mining~\cite{UiPathPM}.

The Process Mining Methodology project aims to improve process performance or compliance to rules and regulations. This typical methodology consists of six phases~\cite{PM2}:
\begin{itemize}
\item
  \textbf{Planning}: During this phase, the business process is selected, the research questions are identified, and the project team is created.
\item
  \textbf{Extraction}: This phase involves determining the scope of the data extraction, based on which event data is to be created. Then, the event data is extracted, and the process knowledge is transferred. 
\item
  \textbf{Data Processing}: In this phase, the views are created, events are aggregated, and event logs are enriched and filtered.       
\item
  \textbf{Mining and Analysis}: In this phase, the process mining techniques and tools are applied on the event logs with the aim of answering the research questions and gaining insights into processes performance and compliance.
\item
  \textbf{Evaluation}: During this phase, the process analysis findings are evaluated to answer the research questions and satisfy the project goals.
\item
  \textbf{Process Improvement and Support}: During this last phase, the goal is to use the insights from the analysis to redesign or improve the current process.
\end{itemize}

Several literature reviews on PM exist, including those of \citet{GARCIA2019} and \citet{PMdrift2022}, but, to our knowledge, none discusses proper integration with RPA. A recent publication from \citet{RPM2022} presents a vision for robotic process mining, supported by many references, but does not result from a systematic review approach. Similarly, \citet{Aalst2021-RPA-PM} present an interesting overview of PM and RPA, but without formally covering the existing literature. Note also that this latter paper is not included in our review as it does not meet our inclusion criteria (i.e., peer-reviewed conference or journal paper).

\subsection{Overview of Robotic Process Automation}
\label{sect:overviewRPA}
Robotic process automation is a fast-emerging process automation approach that uses software robots to replicate human tasks~\cite{Aalst2018}. After recording a process (or workflow) and its steps, a virtual bot is created to mimic the actions performed by humans in that process. RPA is mainly focused on the automation of repetitive, routine, rule-based human tasks, aiming to improve current running processes in an organization.

RPA is defined as an umbrella of tools that operate at the user-interface level of applications the same way humans do~\cite{Aalst2018}. RPA tools perform conditional statements on structured data, typically using a combination of user interface interactions, or by connecting to APIs to drive client servers, mainframes, or Web-based code. \citet{Jimenez2019} highlight the various phases of an RPA project lifecycle, where the starting point is the process assessment to identify candidates for automation, followed by the design and programming of the robots, the implementation/deployment of these robots in their environments, and their maintenance. In general, RPA projects indeed have four distinct stages:

\begin{enumerate}
\def\labelenumi{\arabic{enumi})}
\item
  \textbf{Assess}: before starting any RPA project, it is essential to understand the existing processes that can be automated, and the steps executed in such a process. This stage includes analyzing the context to determine which processes or parts thereof can be automated using RPA technology. This stage also includes the understanding of the design of the selected processes, which involves the specification of the events, data flows, and sequences that must be developed.
\item
  \textbf{Program and test:} during this stage, the discovered processes are turned into RPA scripts that configure the software robots to perform those processes. Testing is performed on each robot to analyze the behavior of the configured bot and detect potential errors that might occur.
\item
  \textbf{Implement}: after completion of the testing stage, the robots are deployed in the production environment to start executing the day-to-day activities of the automated processes.
\item
  \textbf{Monitor and sustain}: once the robots are deployed, it is essential to monitor their performance in case of errors caused by a change in a process step or by a condition not being triggered
  properly. Continuous monitoring can help in performance improvement of the bots and in minimizing the number of errors.
\end{enumerate}

According to Gartner~\cite{Gartner2022}, the RPA software market continues to grow, from 1.26 billion USD in 2020 to 1.61 billion USD in 2021, and with a projected 2.9 billion USD in 2022. Worldwide, the RPA end-user spending is expected to go up by another 17.5\% in 2023. Most of the RPA deployments are industry-specific in the financial and administration sectors. Some available tools in the RPA market include solutions from Automation Anywhere~\cite{AutomationAnywhere}, BluePrism~\cite{BluePrism}, Nintex~\cite{Nintex}, UiPath~\cite{UiPath}, and WorkFusion~\cite{WorkFusion}.

\subsection{Related Literature Reviews on RPA}
\label{sect:relatedRPAreviews}

We note six interesting literature reviews related to RPA since 2019, and they are briefly discussed here from the oldest one to the most recent one. 

In their SLR, \citet{RPASLR2019} studied the state of the art of RPA, including trends and applications. Their research questions focused on the progress of RPA as research field, the difference between RPA and Business Process Management Systems, and the use of RPA in business practice.

The SLR of \citet{Beetz2019RoboticPA} reviews papers that focus on process evaluation development to identify the business processes that are suitable for RPA implementations. In the RPA process evaluation model and as part of the pre-selection phase, PM is mentioned as an approach but the paper does not include details on any particular PM techniques. The authors highlighted that special attention must be paid to the progress in PM as this approach is promising to support selecting processes that can be good candidates for RPA implementation. 

In their work, \citet{Syed2020RoboticPA} reviewed 125 papers on RPA. Their six research questions focused on RPA definitions, benefits, readiness, capabilities, methodologies, and technologies. Even though the authors mentioned that PM is a potential approach to help in developing process scripts for RPA bots, they did explore the papers intersecting PM and RPA. 

The literature mapping of \citet{RPASLR2020} looked into 54 primary studies on RPA's state of the art. Its main conclusion is that the Assess phase of the RPA project lifecycle is not properly covered in most commercial tools. These tools lack proper features for assessing which candidate processes can be automated using RPA.

The SLR of \citet{BPMRPA2021} focuses on the interrelations between process management and RPA from a managerial perspective. It highlights that RPA implementations change the nature of processes, though such implementations come with many challenges, and that RPA encourages redesigning and improving processes.

\citet{CostaRPASLR2022} focused on the RPA adoption aspect in their SLR. The main purpose of their work is to identify the general implementation aspects for a successful RPA adoption, including benefits, challenges, and making processes suitable for RPA. A first outcome of this SLR is providing a way for organizations to have a better understanding of good practices to adopt RPA. A second outcome is highlighting the importance to do further research on this topic.

Note that none of these literature reviews properly addresses the intersection of PM and RPA, while several reviews identify PM as a promising venue for process selection in an automation context that deserves further research. This justifies in part the focus of our own review.

\subsection{Overview of Task Mining}
\label{sect:TaskMining}
Task mining is a collection of techniques for discovering and analyzing the execution of tasks done by humans, based on records of interactions between humans/workers and software applications~\cite{Pareto-DATA2020}. Using task mining, the interactions of workers with their workstations can be recorded and analyzed as they are linked to a process. The three main use cases for task mining are: task discovery and optimization, resource and workforce optimization, and task automation. The recent literature review of \cite{Mayr2022taskMining} provides additional insights on task mining, with a discussion of important challenges at that level, including the collection, protection, explainability, combination, processing, and segmentation of relevant data.

\subsection{An Illustrative Example}
\label{sect:illustrativeexample}
\begin{figure}
    \centering
    \begin{subfigure}{.32\textwidth}
      \centering
      \includegraphics[width=\linewidth]{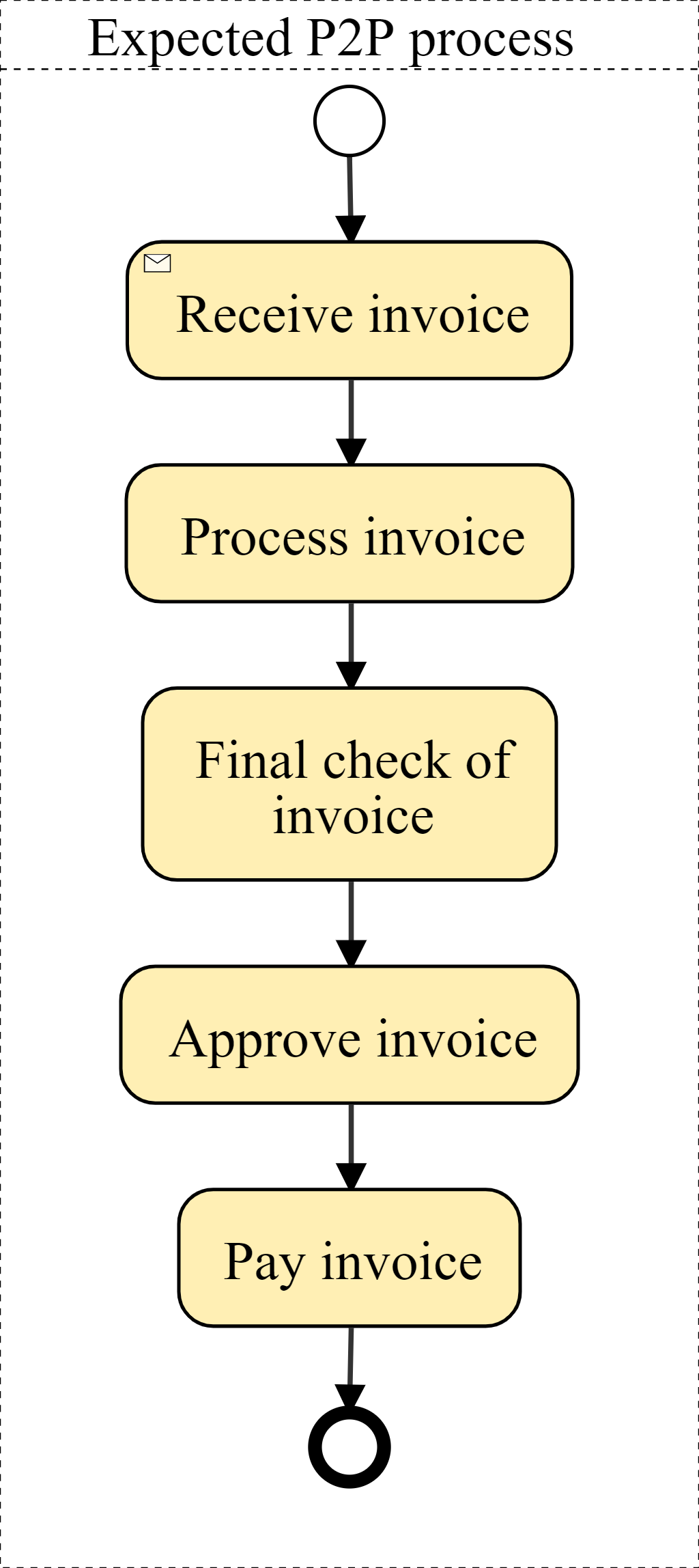}
      \caption{Expected P2P process}
      \label{fig:p2p-expected}
    \end{subfigure}%
    \begin{subfigure}{.68\textwidth}
      \centering
      \includegraphics[width=0.95\linewidth]{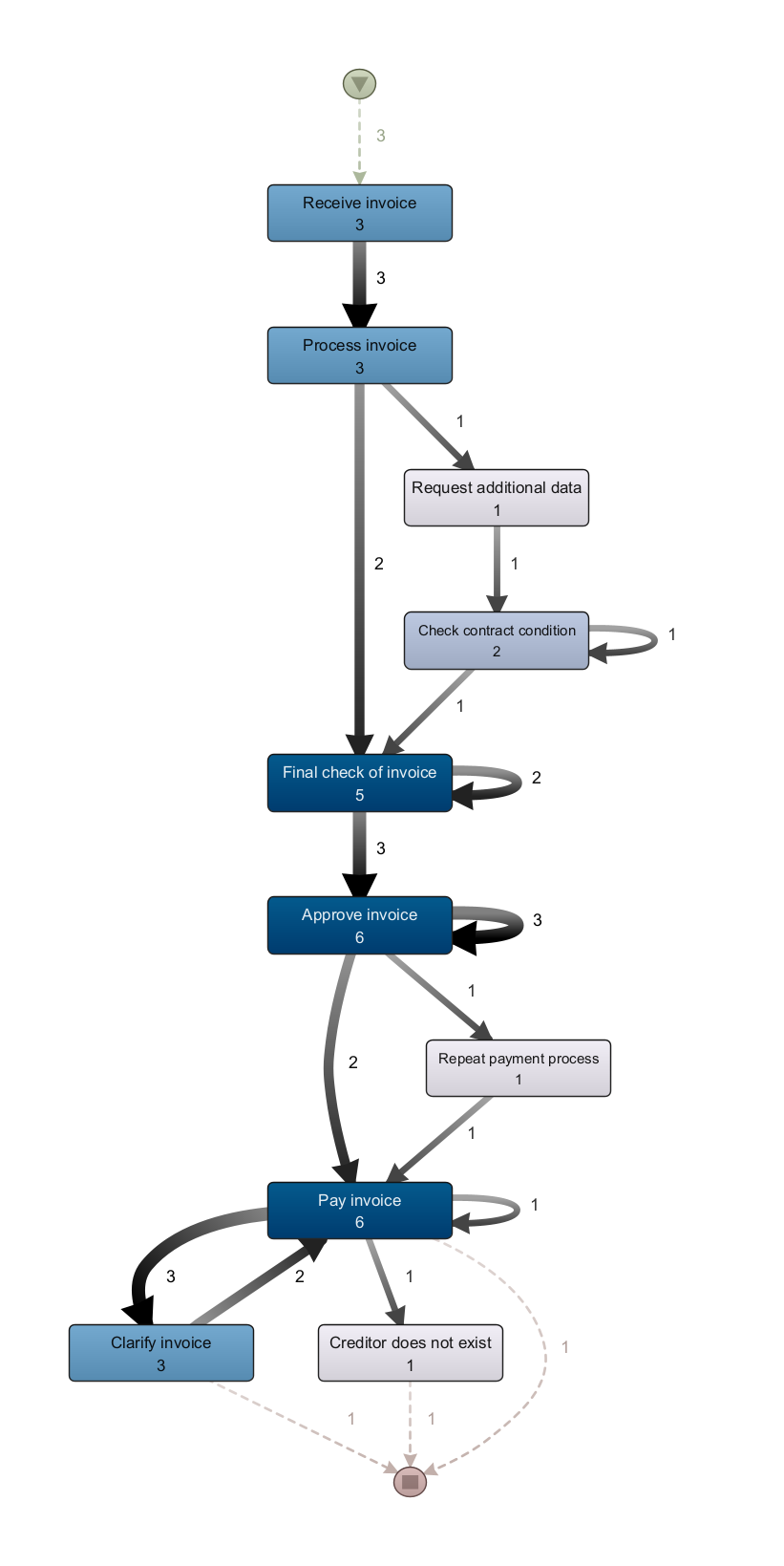}
      \caption{As-is P2P process (generated by Disco)}
      \label{fig:p2p-DISCO}
    \end{subfigure}
    \caption{Procure-to-Pay (P2P) process}
    \label{fig:p2p}
\end{figure}

As discussed in Section~\ref{sect:introduction}, process mining can potentially accelerate and improve the implementation of robotic process automation. For example, suppose that an organization wants to create RPA robots for their \emph{Procure-to-Pay} (P2P) process. P2P is a core process for most companies because it drives value and profitability in the business. The organization believes that its P2P process consists of five different steps that are executed in the order presented by the model in Figure~\ref{fig:p2p-expected}, expressed in the Business Process Model and Notation (BPMN) syntax, commonly used in process maps/models.

After extracting an event log from their information systems, the organization analysts use process mining to model their real as-is process. The event log dataset includes information about the transactions related to the P2P process. The main attributes in the event log are transaction number, timestamp, event, department, cost, and items orders. As shown in Figure~\ref{fig:p2p-DISCO}, the as-is process produced by the Disco PM tool is quite different from the expected model. All the steps executed in the as-is process need to be taken into condition when developing the software robots, otherwise robots will fail to execute a step if a condition is not met. Process mining discovery algorithms and tools help analyze and understand the as-is process steps and eliminate the guess work. Using PM tools, the organization can detect long purchasing cycle times due to bottlenecks in the purchase requisition and order process, supplier delivery delays, and long cycle time to process invoices due to incomplete or missing data.One important challenge in using PM here is that when used on routines that involved fined-grained software-level tasks (e.g., copy-pasting from one application to another one), the complexity of the processes can grow very quickly unless proper abstraction is used.

Another benefit of process mining in this example is the identification of steps that should be automated to improve the efficiency of a process. The Disco tool reports that it takes 7 days to perform \textsf{Process invoice} from the time the invoice is received, perhaps because the staff are busy or because of issues faced along the way. This is a step that can be automated to reduce the time it takes to execute this particular process.

The as-is process in Figure~\ref{fig:p2p-DISCO} also highlights \emph{loops} in several parts of the process, e.g., for \textsf{Final check of invoice}, \textsf{Approve invoice}, and \textsf{Pay invoice}. The discovered process raises important questions, such as: 1)~\emph{Why are the \textsf{Approve invoice} and \textsf{Pay invoice} steps repeated several times?} and 2)~\emph{What are the issues that caused this to happen?} These could be deviations from the expected path that should be prevented. In such cases, creating software robots to execute the process steps can help improve the quality of a process and eliminate unnecessary steps or repetitions while minimizing errors.

After discovering the as-is process, starting an RPA project implementation involves programming a robot to execute the automatable tasks, e.g., from the Receive invoice step to the Process invoice step. The robot is programmed to take the inputs from the Receive invoice application and input them into the required fields so the invoice can be processed. In the \textsf{Process invoice} step of the P2P process, the related tasks can be automated using RPA: manage invoice collection and entry, manage electronic invoicing, validate, and handle invoice data, and submit transactions for processing. The bot will be programmed to:

\begin{enumerate}
\def\labelenumi{\arabic{enumi})}
\item
  Scan a mailbox for orders
\item
  Log into the necessary systems
\item
  Read the invoice image
\item
  Register the invoice in the SAP system
\item
  Perform all the necessary validations, including cross-checking against other systems, and decide whether to post, park, or block the invoice.
\end{enumerate}

The process can be optimized through automation using the knowledge gained in the PM discovery phase., e.g., to evaluate the supplier performance and share results to create an improvement plan with quantifiable goals. The robots can be programmed to handle purchase order approvals up to a certain threshold and ask humans to handle approvals above that threshold. The robots can run the payment process to ensure payment happens at the right time, which will result in higher accuracy and faster invoice processing time. The software bots can handle high-volume repetitive tasks, reduce errors, improve performance, and enhance productivity.

After the RPA robots are deployed and start executing the different steps, \emph{another} process map can be generated to visualize the improvement in the process. The organization could hence check whether the average time has been reduced to a few minutes from \textsf{Receive invoice} to \textsf{Process invoice}. The robots are hence monitored using process mining techniques, which in turn enables the organization to monitor purchase order cycle times, supplier lead times, delivery dates, and invoice processing times. Such knowledge allows the organization to act in order to keep its processes on track. The organization can set up alerts to inform managers when 1)~a supplier delivers late, 2)~a purchase needs managers approval, 3)~a payment is at risk of being delayed, and 4)~there is missing information at any step in the process.

From this example, it is important to observe that process mining can help make appropriate decisions regarding what processes and parts thereof to automate via RPA, and monitor the effectiveness of the RPA-based automation.

\section{Literature Review Methodology}
\label{sect:methodology}
This section presents the methodology used to select and review the literature related to how PM techniques are used to accelerate and improve the implementation of RPA projects. The main goal in conducting this review is to provide a summary of the current literature to answer the research questions. The studies focusing on the combination of process mining techniques in the early stage to build robotic process automations are discussed. The systematic literature review was undertaken based on the guidelines of Kitchenham and Charters (2007). The methodology for this SLR, research questions, research query, inclusion/exclusion criteria, and analysis criteria used to answer the research question are presented here, together with the list of selected papers.

\subsection{Research Questions}
\label{sect:RQs}
The introduction has presented three research questions this paper aims to answer, through the lens provided by the literature. The first question (RQ1) deals with identifying the different studies that use PM techniques and algorithms to discover processes that can be automated using RPA. The second question (RQ2) aims to highlight both PM and RPA tools that have been used in the selected studies, and whether they are open source or commercial tools. The third question (RQ3) aims to list the challenges that exist when combing PM and RPA together, and whether these challenges are mainly about the PM part, the RPA part, or both parts.

\subsection{Search Process and Query}
\label{sect:query}

In this research, seven of the most relevant search engines in information technologies were used. Elsevier's Scopus and Clarivate's Web of Science are both wide-range, curated, general databases. Web of Science is useful to cover older papers, whereas Scopus indexes more conference and journal papers. In addition, the engines of popular publishers in the PM and RPA areas were included, to consider more recent work not covered by the two previous engines: IEEE Xplore, ACM Digital Library, SpringerLink, and Elsevier's ScienceDirect. arXiv was also added to capture more recent papers (at times from unreliable sources) that were not indexed by the other databases.

Since this SLR focuses on the papers that address PM and RPA, the main two concepts are \emph{``process mining''} and \emph{``robotic process automation''}, and their common synonyms such as \emph{``task mining''}, \emph{``segmentation''}, \emph{``user interaction''}, and \emph{``UI logs''}. The main query (last searched in November 2022) is the following:

\vspace{0.25cm}
\textsf{
("process mining" OR "process discovery")\\ 
AND \\
\indent("robotic process automat*" OR "intelligent process automat*" OR RPA\\
\indent ~~OR "segmentat*" OR "UI log*" OR "user inter* log*" OR "task mining") 
}
\vspace{0.25cm}

No specific time limit was used. A total of 605 unique papers (767 including duplicates) were returned. Table~\ref{tab:databases} details the number of papers returned by each of the different databases. The number of papers returned is reasonable considering that this is a new topic in research and there have  been an increasing number of publications at the intersection of PM and RPA. SpringerLink returned a higher number of papers (296) because it searches the keywords in the whole text of the papers and not just in the title and abstract. arXiv had the same issue. Consequently, many papers were returned that were not relevant to answer the research questions. As advised by \citet{MOURAO2020-Snowballing}, a one-level snowballing approach exploiting the papers' references was used on the selected papers in order to gather more papers possibly relevant to the topic but that were not captured by the database searches. Only \textit{backward} snowballing (i.e., towards older references contained in the selected papers) was used.

\begin{table}[]
    \centering
    \small
    \begin{tabular}{l|l|l}
        \hline
        \textbf{Database} & \textbf{Searched Within} & \textbf{Papers Returned} \\
        \hline
        Scopus & Title/abstract/keywords & 89 \\
        IEEE Xplore & Title/abstract/keywords & 167 \\
        Web of Science & Title/abstract/keywords & 36 \\
        ACM DL & Title/abstract/keywords & 12 \\
        SpringerLink & All Text & 296 \\
        ScienceDirect & Title/abstract/keywords & 117 \\
        arXiv & All Fields & 50 \\
        \hline
    \end{tabular}
    \caption{Number of Papers Returned by the Search Engines}
    \label{tab:databases}
\end{table}

\subsection{Inclusion and Exclusion Criteria}
\label{sect:criteria}
In order to exclude the papers that are irrelevant to our research questions, exclusion criteria were defined. These criteria were used to exclude papers that:

\begin{itemize}
\item Do not focus on how PM or process discovery algorithms are used to better build RPA.
\item Only refer to how to apply PM on RPA event logs (as RPA was already implemented).
\item Were not written in English.
\item Were published in predatory journals or conferences.
\end{itemize}
One example from \citet{botProcessMining} applied process mining techniques on logs collected from RPA bots; such papers were excluded from the list of selected papers because of the second criterion. The inclusion criteria were:

\begin{itemize}
\item Peer-reviewed conference proceedings, articles, and journal publications.
\item Publications added through the one-level backward snowballing strategy.
\end{itemize}

\subsection{Paper Selection}
The results of the database searches were imported into Covidence\footnote{\url{https://www.covidence.org/}}. Duplicate results were identified by Covidence and eliminated. Then, the title and abstract for each of the papers were screened to eliminate the obvious irrelevant studies.

After scanning the papers returned by the search with the inclusion and exclusion criteria, the full text of all the selected papers were retrieved. Snowballing on the references of the selected papers (based on the references they contained and on papers referencing them on Google Scholar) resulted in the discovery of three more relevant papers.

\begin{figure}
    \centering
    \includegraphics[width=\textwidth]{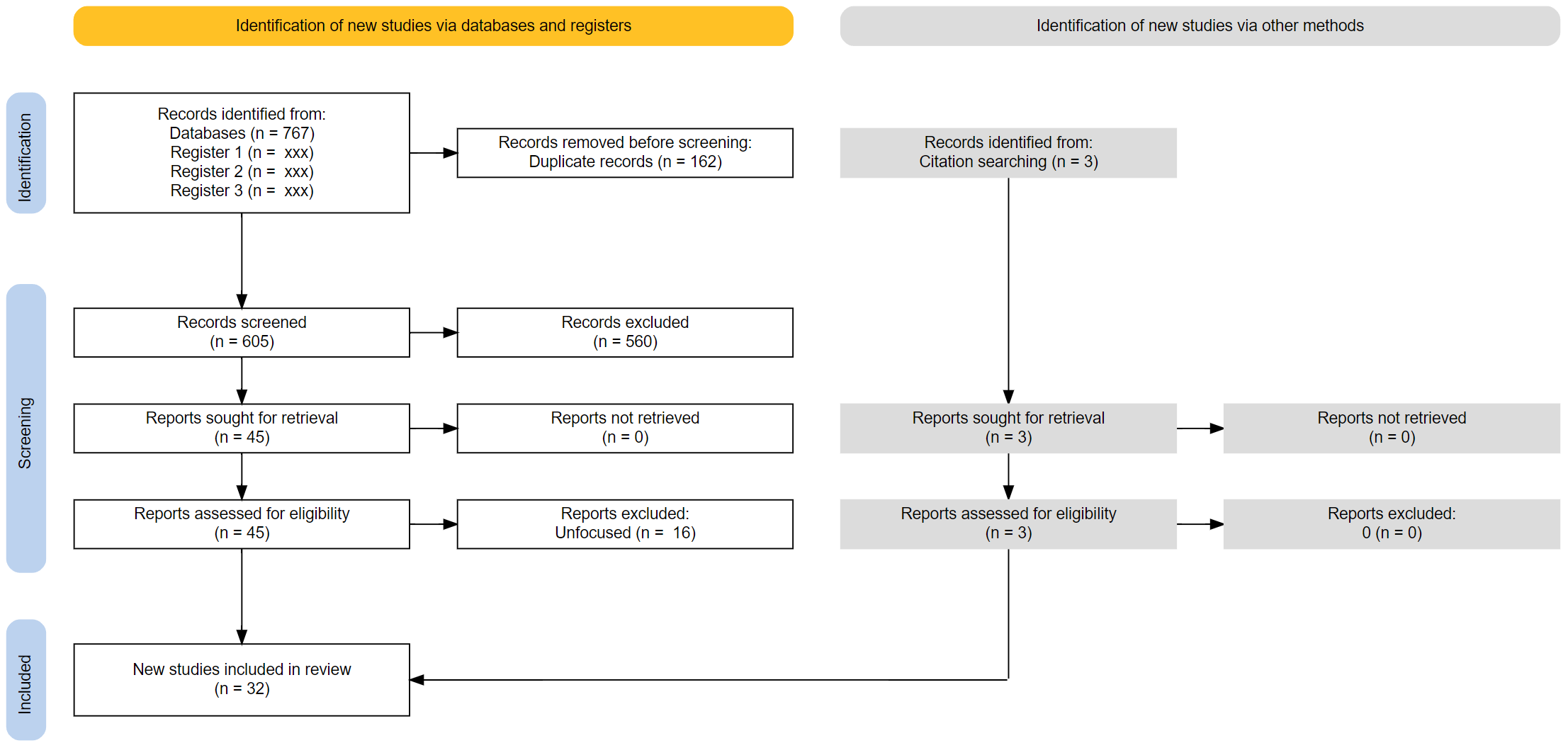}
    \caption{PRISMA 2020 flow diagram for this SLR}
    \label{fig:PRISMA}
\end{figure}

Figure~\ref{fig:PRISMA} shows the PRISMA diagram\footnote{\url{http://prisma-statement.org/prismastatement/flowdiagram.aspx}} for the number of papers that were included in this SLR at each stage of the search methodology. The next step was reading the papers and extracting the data. The extracted data includes details about each paper including the title, date, year of publication, type, objective, algorithm developed, challenges, and tools used.

\subsection{Final Selection}
\label{sect:selection}
A total of 32 papers, listed in Table~\ref{tab:papers}, were selected for this SLR. All returned papers were published in or after 2018, with the largest number of publications in 2022. Note that no paper was selected from arXiv in the end, and thus the selected papers are all peer-reviewed.

\begin{table}[]
    \centering
    \resizebox{\textwidth}{!}{
    \begin{tabular}{l|l|l}
    \hline
    \textbf{Paper} & \textbf{Title} & \textbf{Year} \\
    \hline
    \citet{Geyer-Klingeberg2018124} & Process Mining and Robotic Process Automation a Perfect Match & 2018 \\
    \citet{Leno2018MultiPerspectivePM} & Multi-Perspective Process Model Discovery for Robotic Process Automation & 2018 \\
    \citet{Linn2018DesktopAM} & Desktop Activity Mining -- A new level of details in mining business processes & 2018 \\
    \citet{Agostinelli2019ResearchCF} & Research Challenges for Intelligent Robotic Process Automation & 2019 \\
    \citet{Bosco2019} & Discovering Automatable Routines from User Interaction Logs & 2019 \\
    \citet{Jimenez2019} & A Method to Improve the Early Stages of the Robotic Process Automation Lifecycle & 2019 \\
    \citet{Kirchmer2019} & Value-Driven Robotic Process Automation (RPA): A Process-Led Approach to Fast Results at Minimal Risk & 2019 \\
    \citet{LenoV2019AlEp} & Action Logger: Enabling Process Mining for Robotic Process Automation & 2019 \\
    \citet{wanner2019process}& Process Selection in RPA Projects -- Towards a Quantifiable Method of Decision Making & 2019 \\
    \citet{Agostinelli2020} & Automated Generation of Executable RPA Scripts from User Interface Logs & 2020 \\
    \citet{Cabello2020} & Beyond the Hype: RPA Horizon for Robot-Human Interaction & 2020 \\
    \citet{Halaska2020} & Importance of Process Flow and Logic Criteria for RPA Implementation & 2020 \\
    \citet{Leno2020RoboticPM} & Robotic Process Mining: Vision and Challenges & 2020 \\
    \citet{Leno2020IdentifyingCR} & Identifying Candidate Routines for Robotic Process Automation from Unsegmented UI Logs & 2020 \\
    \citet{Pareto-DATA2020} & On the Pareto Principle in Process Mining, Task Mining, and Robotic Process Automation & 2020 \\
    \citet{Agostinelle2021} & Exploring the Challenges of Automated Segmentation in RPA & 2021 \\
    \citet{choi2021} & Candidate Digital Tasks Selection Methodology for Automation with Robotic Process Automation & 2021 \\
    \citet{rinderle2021} & Process Automation and Process Mining in Manufacturing & 2021 \\
    \citet{LENO2022101916} & Discovering Data Transfer Routines from User Interaction logs & 2021 \\
    \citet{luka2022} & A Reference Data Model for Process-related User Interaction Logs & 2022 \\
    \citet{AMartinesRojas2022} & Analyzing Variable Human Actions for Robotic Process Automation & 2022 \\
    \citet{FabianRPAFlowcharts2022} & Process Discovery Analysis for Generating RPA Flowcharts & 2022 \\
    \citet{humaninLoop2022} & A Human-in-the-loop Approach to Support the Segments Compliance Analysis & 2022 \\
    \citet{Manufacture2022} & Intelligent Process Automation: An Application in Manufacturing Industry & 2022 \\
    \citet{Cabello2022} & Hybridizing Humans and Robots: An RPA Horizon Envisaged from the Trenches & 2022 \\
    \citet{Neelam2022} & A Path Forward for Automation in Robotic Process Automation Projects: Potential Process Selection Strategies & 2022 \\
    \citet{Choi2022} & Enabling the Gap Between RPA and Process Mining: User Interface Interactions Recorder & 2022 \\
    \citet{HanFinancialRobots2022} & A Review on Financial Robot Process Auto-mining Based on Reinforcement Learning & 2022 \\
    \citet{PMandRPAP2P} & Improving Purchase to Pay Process Efficiency with RPA using Fuzzy Miner Algorithm in Process Mining & 2022 \\
    \citet{RPM2022} & Robotic Process Mining & 2022\\
    \citet{AGOSTINELLI2022} & Reactive Synthesis of Software Robots in RPA from User Interface Logs & 2022\\
    \citet{AgostinelliInteractiveUI2021} & Interactive Segmentation of User Interface Logs & 2022\\
    \hline
    \end{tabular}
    }
    \caption{Papers selected for the SLR, sorted by year of publication}
    \label{tab:papers}
\end{table}

\subsection{Analysis}
From each selected paper, we extracted:
\begin{itemize}
    \item 
        The title, authors, and year of publication (as commonly collected metadata).
    \item
        The algorithms and techniques used for PM and for RPA, with informal summaries, to answer RQ1. Each of them was also manually mapped (from the text) to one or many of the eleven steps described in Figure~\ref{fig:steps}.
    \item 
        The PM and RPA tools used, to answer RQ2.
    \item 
        The identifiable PM-related challenges and RPA-related challenges, to answer RQ3.
\end{itemize}

The next section reports on the results synthesis from that analyzed information.

\section{Results of Selected Studies}
\label{sect:results}
After reviewing and analyzing the selected papers, and in order to answer the research questions for this SLR, three themes are defined: techniques and algorithms (RQ1, covered in Section~\ref{sect:techs-algos}), tools (RQ2, in Section~\ref{sect:tools}), and challenges (RQ3, in Section~\ref{sect:challenges}).

\begin{figure}
    \centering
    \includegraphics[width=5.4in]{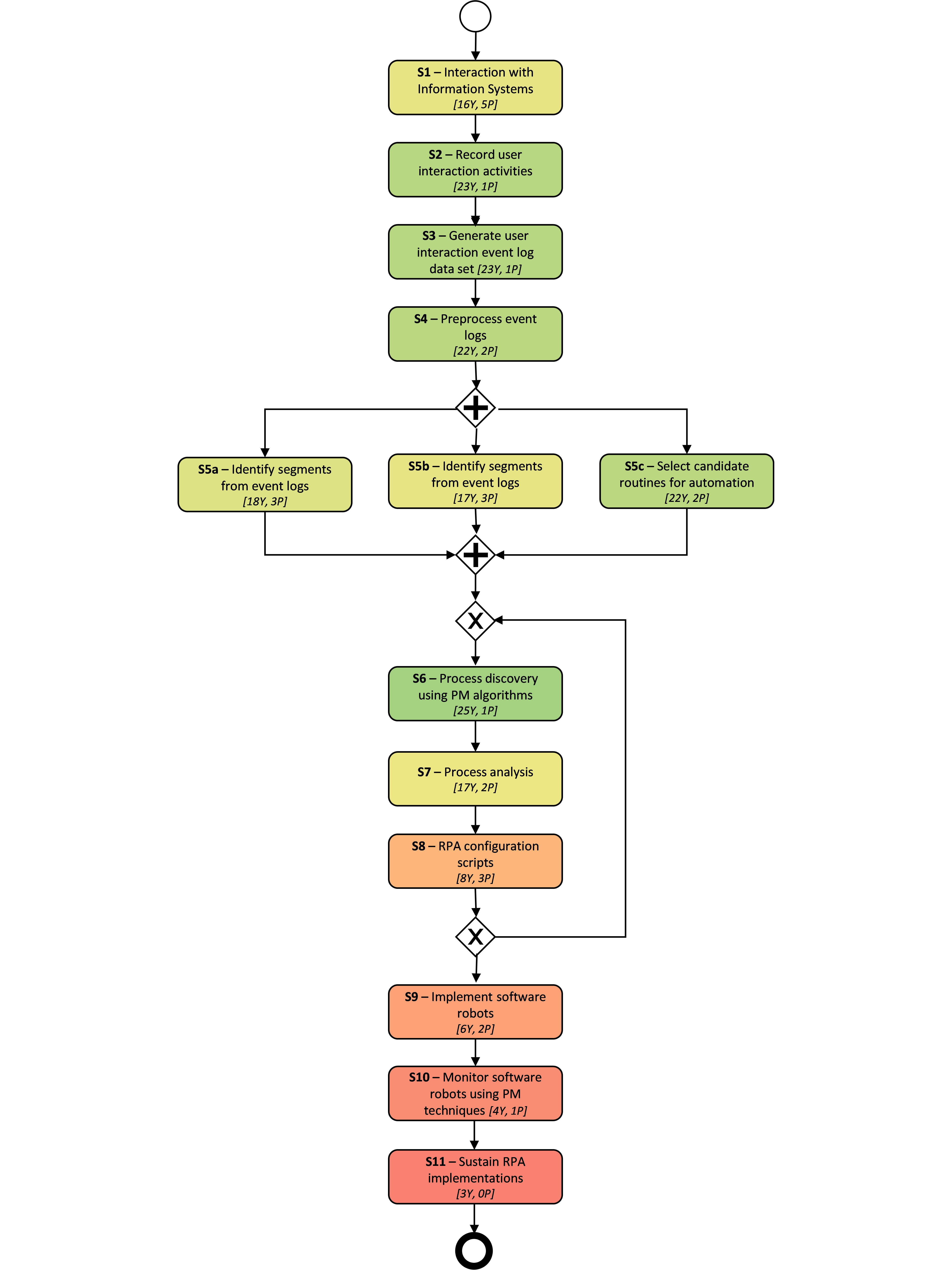}
    \caption{Typical PM+RPA process steps, with frequency among the 32 selected papers [\#yes, \#partial]. The step colors reflect their frequencies (the greener, the higher).}
    \label{fig:steps}
\end{figure}

Figure~\ref{fig:steps} illustrates, using the BPMN syntax, the steps from a successful PM+RPA project lifecycle that involves process mining techniques and algorithms to implement RPA solutions. The steps S1 to S11 will be used to assess the coverage of existing techniques. These steps were identified iteratively based on the papers that were reviewed as part of this literature. Each step also reflects, between square brackets, how many of the 32 papers cover that step fully (Y) or partially (P). These frequencies are combined into a heatmap representation, where redder colors are closer to a 0 frequency, greener colors are closer to 31, and yellow is used in between (16). This heatmap already shows that the earlier steps are far better covered by the literature than the latter steps. More details are provided in this section, especially around Table~\ref{tab:papers-analysis}. 

\subsection{Techniques and Algorithms}
\label{sect:techs-algos}
This section reports on the techniques and algorithms highlighted in each paper according to the steps they cover in Figure~\ref{fig:steps}. Section~\ref{subsect:collecting} focuses on collecting interactions with various information systems (S1) whereas Section~\ref{subsect:recording} covers recording activities and the generation of event logs (S2-S3). Section~\ref{subsect:preprocessing} then 
discusses the preprocessing of event logs (S4-S5), whereas Section~\ref{subsect:discovery-analysis} focuses on discovery and analysis activities (S6-S7) and, finally, Section~\ref{subsect:RPA-config} covers RPA configuration (S8) and sustainability (S9-S11). Section~\ref{subsect:PM-RPA-Start2End} discusses the papers that cover the most steps from start (S1) to end (S11).

\subsubsection{Collecting Interactions with Information Systems}
\label{subsect:collecting}
The starting point of process mining is the existence of event logs that include information about the processes that were executed. Step S1 in Figure~\ref{fig:steps} refers to the interaction and connection with information systems to collect event logs. However, currently most of these event
logs are collected by information systems and do not include many details about actions done by users at the user interactions (UI) level.

In order to implement a successful RPA project and minimize the number of errors after deploying software robots, it is essential to understand the details of a process at a UI level since these are the steps that will be automated and executed using software robots. Several papers, discussed in the next section, focus on proposing new techniques to collect more logs at the user interaction level of an application then applying PM algorithms on the collected logs to generate process maps and discover the process steps. 

\subsubsection{Recording Activities and Generating Event Logs} 
\label{subsect:recording}
Step S2 in Figure~\ref{fig:steps} is concerned with recording user interaction activities. During this step, the user interactions with applications are recorded at a very granular level, e.g., \emph{copy} an Excel cell with its value, and \emph{paste} actions. Usually, process mining techniques have been applied on event logs collected from information system that do not include this level of detail. However, in the context of RPA, it is essential to record user interaction activities. Current PM tools do not generally provide features for collecting such types of activities, whereas some RPA tools provide \emph{task mining} and \emph{task capturing} features. For example, the UiPath Task Mining tool~\cite{UiPathTM} automatically identifies the employee workflow, and then identifies repetitive tasks that can be automated. Thus, several papers focus on recording user interaction activities that include collecting events and actions observable at the UI level.

\citet{Linn2018DesktopAM} introduced the concept of Desktop Activity Mining (DAM), which is designed to add a new level of detail in process mining by a)~capturing user actions that are not logged by information systems, b)~obtaining a complete set of process variants, and c)~deriving a process model and documentation about the process details. 
DAM is a method to record user activities at the level of desktop actions and to reconstruct the resulting process variations with process and data mining techniques to discover a process model. \citet{Jimenez2019} proposed a method to improve the early stages of RPA project lifecycles. Their approach is to monitor back-office staff through a screen-mouse-key-logger, with the obtained event logs transformed to a UI log used as an input to PM algorithms. This method reduces the effort in analyzing the actual system to discover the processes that can be automated manually, rather than using process mining techniques.

\citet{Leno2018MultiPerspectivePM} moved a step further in the process and worked on generating process maps. The authors particularly look at multi-perspective process model discovery for robotic process automation. In their research project, they aim to develop process mining technology to extract a flowchart of how users interact with a given UI, which can then be used to train and test RPA bots automatically. This research is expected to accelerate the adoption of RPA solutions. Process mining could be a way to solve the problem of discovering such flowcharts, but the user interactions logs are not typically collected by information systems and are not fully documented. Thus, the process mining approach does not have access to the activities performed in such systems and such event logs should be collected explicitly.

\citet{LenoV2019AlEp} introduced a new tool called Action Logger for recording UI logs, which are logs of the user interactions with information systems. The purpose of the tool is to collect UI logs at a granular level suitable to discover the sufficient details of a process steps in the context of RPA. This paper focuses on steps S1 to S3 to collect the sufficient event logs needed. Its tool includes functional requirements not considered in other papers that aim to collect more UI logs. Action Logger has five functional requirements for collecting logs, as summarized in Table~\ref{tab:ActionLogger}.

\begin{table}[]
    \small
    \centering
    \begin{tabular}{p{.35\textwidth}p{.57\textwidth}}
    \hline
    \textbf{Functional Requirement} & \textbf{Definition} \\
    \hline
    Relevance & Record meaningful actions, for example moving the mouse
    action should not be recorded whereas copying a cell should be. \\
    Granularity & Record actions at a level of detail needed to understand
    the tasks that were performed. \\
    Data-awareness & Record the data that supports each of the actions that
    are recorded, for example, recording the timestamp of a particular
    action. \\
    Context-independence & Record actions with information about the
    platforms where the actions were performed, with their various
    circumstances and context. \\
    Interoperability & Record actions and logs in a format that can be
    supported by process mining tools. \\
    \hline
    \end{tabular}
    \caption{Action Logger functional requirements}
    \label{tab:ActionLogger}
\end{table}

The Action Logger tool architecture only records UI events happening in Excel and Chrome using an Excel plug-in and a Chrome plug-in. The two plug-ins are implemented as event listeners and send the information about the performed actions as JSON objects to a logging component that generates the UI log. \citet{LenoV2019AlEp} did not discuss how many users are executing the tasks and how this is being recorded with the Action Logger. Their tool did not include any steps on how to select the paths that are suitable for automation and the authors did not discuss challenges encountered at this level. Additionally, \citet{LenoV2019AlEp} mentioned briefly that they developed a log simplification tool to reduce the size of UI logs before importing them into PM tools (Apromore in their case), which was implemented as set of regular-expression search-and-replace rules\footnote{Log Simplification tool available at \url{https://github.com/apromore/RPA\_SemFilter/releases}}.

\citet{Agostinelli2020} developed the SmartRPA tool~\cite{SmartRPA}, which generates executable RPA scripts that can be automated using software robots by exploiting UI event logs. SmartRPA skips the manual work required to discover flowcharts by discovering process maps from UI logs recorded from the user interacting with the software application. SmartRPA consists of five stages: 1)~record UI logs for the different routine executions, 2)~combine UI logs into a single event log, 3)~filter out irrelevant routines, 4)~detect the most frequent routine variant for a process from the event log, and 5)~generate the executable RPA scripts to create the software robots. The SmartRPA tool was developed in Python.

Additionally, SmartRPA records only those UI actions that can be automated and can be associated with routines. It also enables recording a large set of UI actions, not just limited to Excel and Chrome, which goes beyond what was developed in other approaches~\cite{Bosco2019,LenoV2019AlEp}.

SmartRPA focuses on steps S1 to S3 of Figure~\ref{fig:steps} and automates the best processes in terms of the frequency and time duration recorded for routine variants, without requiring an a-priori model. SmartRPA records events that happen during a UI interaction, so it can work across different computer systems. The identification of similar routine variants is not done using the screens of the user's desktop, which may differ between different computer systems.

\citet{AgostinelliInteractiveUI2021} presented an approach to discover the routine traces from unsegmented UI logs, which depends on:
\begin{enumerate}
\item
  A frequent pattern identification technique to automatically derive routines as recorded in UI logs.
\item
 A human-in-loop interaction to filter out those segments not allowed by real-world routines under analysis.
\item
  A trace alignment technique to cluster all those user actions belonging to specific segments into routine traces.
\end{enumerate}

\begin{table}[]
    \centering
    \resizebox{\textwidth}{!}{
    \begin{tabular}{l|l|l|l|l|l|l|l|l|l|l|l|l|l}
    \hline
    \textbf{Paper} & \rotl{S1: IS interaction} & \rotl{S2: recording} & \rotl{S3: generation} & \rotl{S4: preprocessing} & \rotl{S5a: segmentation} & \rotl{S5b: routines} & \rotl{S5c: select} & \rotl{S6: discovery} & \rotl{S7: analysis} & \rotl{S8: RPA config} & \rotl{S9: robot impl.} & \rotl{S10: monitoring} & \rotl{S11: sustain} \\
    \hline
    \citet{Geyer-Klingeberg2018124} &\N &\N &\N  &\N &\N  &\N  &\N  & \Y & \Y & \Y & \Y & \Y & \Y   \\
    \citet{Leno2018MultiPerspectivePM} &\Pa & \Y &\Y &\Y &\Pa &\Pa &\Pa &\N &\N &\N &\N &\N &\N   \\
    \citet{Linn2018DesktopAM} &\Pa & \Y &\Y &\Y &\N &\N &\Pa &\Y &\Y &\Y &\N &\N &\N  \\
    \citet{Agostinelli2019ResearchCF} & \Y &\Y &\Y &\Y &\Y &\N &\N &\N &\N &\N &\N &\N &\N \\
    \citet{Bosco2019} &\Pa &\Y &\Y &\Y & \Y & \Y & \Y &\N &\N &\N &\N &\N &\N \\
    \citet{Jimenez2019} &\Y & \Y &\Y &\Y &\Y &\Y &\Y &\Y &\Y &\N &\N &\N &\N \\
    \citet{Kirchmer2019} &\N &\N &\N &\N &\N &\N &\N &\Y &\Y &\Y &\Y &\Y &\Y\\
    \citet{LenoV2019AlEp} &\Y & \Y &\Y &\Y &\Y &\Y &\Y &\Y &\N &\N &\N &\N &\N  \\
    \citet{wanner2019process} &\Y &\Y &\Y &\Pa &\N &\N &\N &\Y &\Y &\N &\N &\N &\N \\
    \citet{Agostinelli2020} &\Y &\Y &\Y &\Y &\Pa &\Pa &\Y &\Y &\Y &\Y &\Y &\N &\N \\
    \citet{Cabello2020}&\N &\Y &\Y &\Y &\N &\N &\N &\Y &\Y &\Pa &\Pa &\Pa &\N\\
    \citet{Halaska2020} &\N &\N &\N &\N &\N &\N &\N &\Y &\Y &\Y &\Y &\Y &\N \\
    \citet{Leno2020RoboticPM} &\Y &\Y &\Y &\Y &\Y &\Y &\Y &\Y &\Pa &\Pa &\N &\N &\N\\
    \citet{Leno2020IdentifyingCR} &\Y &\Y &\Y &\Y & \Y & \Y &\Y &\Y &\N &\N &\N &\N &\N \\
    \citet{Pareto-DATA2020} &\N &\N &\N &\N &\N &\N & \N &\Y &\Y &\N &\N &\N &\N\\
    \citet{Agostinelle2021} &\Pa &\Pa &\Y &\Y &\Y &\Y &\Y &\Y &\N &\N &\N &\N &\N \\
    \citet{choi2021} &  \Y &\Y &\Y &\Y & \Y & \Y & \Y &\Y &\Pa &\N &\N &\N &\N \\
    \citet{rinderle2021} &\Pa &\Y &\Y &\Y &\N &\N &\Y &\Y &\Y &\Y &\N &\N &\N \\
    \citet{LENO2022101916} &\Y &\Y &\Y &\Y & \Y & \Y &\Y &\Y &\N &\N &\N &\N &\N \\
    \citet{luka2022} &\N &\Y &\Y &\Y &\Y &\Y &\Y &\Y &\N &\N &\N &\N &\N \\
    \citet{AMartinesRojas2022} &\Y &\Y &\Y &\Y &\Y &\Y &\Y &\N &\N &\N &\N &\N &\N \\
    \citet{FabianRPAFlowcharts2022} &\N &\Y &\Y &\Y &\Pa &\Pa &\Y &\Y &\Y &\Y &\Y &\N &\N\\
    \citet{humaninLoop2022} &\Y &\Y &\Y &\Y &\Y &\Y &\Y &\Y &\Y &\N &\N &\N &\N \\
    \citet{Manufacture2022} &\N &\N &\N &\N &\N &\N &\Y &\Y &\Y &\N &\N &\N &\N \\
    \citet{Cabello2022} &\Y &\Y &\Y &\Y &\Y &\Y &\Y &\Y &\Y &\N &\N &\N &\N \\
    \citet{Neelam2022} &\Y &\Y &\Y &\Y &\Y &\Y &\Y &\Pa &\N &\N &\N &\N &\N \\
    \citet{Choi2022} &\Y &\Y &\Y &\Y &\Y &\Y &\Y &\Y &\Y &\N &\N &\N &\N \\
    \citet{HanFinancialRobots2022} &\Y &\N &\N &\N &\N &\N &\N &\Y &\Y &\Pa &\Pa &\N &\N \\
    \citet{PMandRPAP2P} &\N &\N &\N &\N &\N &\N &\Y &\Y &\Y &\Y &\Y &\Y &\Y \\
    \citet{RPM2022} &\N &\Y &\N &\N &\Y & \Y &\Y &\N &\N &\N &\N &\N &\N  \\
    \citet{AGOSTINELLI2022} &\Y &\Pa &\Pa &\Pa &\Y &\Y &\Y &\N &\N &\N &\N &\N &\N \\
    \citet{AgostinelliInteractiveUI2021} &\N & \Y &\Y &\Y &\Y &\Y &\Y &\Y &\N &\N &\N &\N &\N\\
    \hline
    \textbf{Frequency \colorbox{green!25}{\checkmark}} & 16 & 23 & 23 & 22 & 18 & 17 & 22 & 25 & 17 & 8 & 6 & 4 & 3\\
    \hline
    \textbf{Frequency \colorbox{yellow!25}{$\pm$}} & 5 & 1 & 1 & 2 & 3 & 3 & 2 & 1 & 2 & 3 & 2 & 1 & 0\\
    \hline    
    \end{tabular}
    }
    \caption{Summary of selected SLR papers, with coverage of PM+RPA steps \newline (\colorbox{green!25}{\checkmark} = Coverage, \colorbox{red!25}{\tiny\strut} = No coverage, and \colorbox{yellow!25}{$\pm$} = Partial coverage).}
    \label{tab:papers-analysis}
\end{table}

An additional tool from \citet{Choi2022} was developed to record interactions with user interfaces and generate UI logs. PM techniques are then applied to those logs to identify the tasks that can be automated using RPA. This approach also focuses on steps S1 to S3 of Figure~\ref{fig:steps}, especially regarding recording, log generation, and log filtering components from the Robotic Process Mining approach developed by \citet{RPM2022}.

During step S3 (generate UI event log datasets), the recorded events are extracted into datasets that can be used for process mining discovery. The dataset needs to include mandatory attributes (columns) for process mining, namely the timestamp, event name, and case ID of each event. Since the recorded events need to be used for generating RPA scripts, the event log dataset will include additional attributes. For instance, in a ``copy cell'' event, an additional attribute specifies the value in the copied cell.

Several vendors, for example Celonis, UiPath, and myInvenio, recently adopted the term \emph{task mining} (see Section~\ref{sect:TaskMining}) to refer to process mining based on UI data. These UI data are collected using task recorders. Often, screen captures are taken to infer actions taken by the user. The challenge is to match UI data based on identifiers, usernames, keywords, and labels, and connect different data sources. Such analysis can be time-consuming to discover all the steps executed by the user involved in the process~\cite{Aalst2018}. 

\citet{luka2022} worked on generating a data model for process-related UI logs. Their approach generates a standardized reference data model for process-related UI logs since often different UI logs rely on different conceptualization and collection techniques. The model was implemented as an extension to XES~\cite{XES} as the standard model for event logs. The authors reviewed both the academic literature and industry tools. They particularly reviewed the core UI attributes such as action type, target element, UI hierarchy, application, input value, and timestamp. These six core attributes were used to assess nine RPA industry tools, and the latter indeed record five of these attributes. The only difference is the support for UI hierarchy, which varies across tools.

\subsubsection{Preprocessing Event Logs}
\label{subsect:preprocessing}
After collecting the raw event log dataset from the UI activities in the applications, the data needs to be prepared for process mining analysis. Given their large size and complexity, the process maps generated from raw data usually do not provide meaningful insights about the steps that are followed to execute a task. During step S4 (Figure~\ref{fig:steps}) for example, the noise events are filtered out, the incomplete process variants are deleted, and the event logs are simplified to provide more meaningful insights when process maps are generated. There has been a research focus on how to preprocess event logs to generate meaning insights when importing them into process mining tools (as discussed by \cite{eventlogprep2021}). Table~\ref{tab:techniques} summarizes the preprocessing techniques at the process and event levels, to be discussed next.

\begin{table}[]
    \centering
    \footnotesize
    \begin{tabular}{p{0.48\textwidth}|p{0.48\textwidth}}
    \hline
    \textbf{Process Level} & \textbf{Event Level}\\
    \hline
    \begin{itemize}
      \item Delete incomplete process variants
      \item Identify segments from event logs
      \item Discover repetitive routines
      \item Decompose UI logs into segments
      \item Unsegment event logs from UI
      \item Generate executable routines
    \end{itemize} 
    &
    \begin{itemize}
      \item Filter out noise events
      \item Mine event patterns 
      \item Discover tasks that can be automated
      \item Restructure events
      \item Transform event names based on action types
    \end{itemize}\\
    \hline
    \end{tabular}
    \caption{Preprocessing techniques at the process and event levels}
    \label{tab:techniques}
\end{table}

Steps S5a, S5b, and S5c are concerned with identifying segments from event logs, identifying candidate routines for automation, and selecting candidate routines for automation, respectively. These three steps focus on preprocessing and restructuring the event logs to discover meaningful routines that can be automated from UI logs, since these types of event logs come with several challenges. The discovery of candidate routines from recorded event logs for automation using RPA solutions is problem seldom explored. Eight papers identified in this SLR focus on using the UI event logs to discover routines that can be automated using RPA \cite{RPM2022, Bosco2019, Leno2020IdentifyingCR, choi2021, LENO2022101916, Neelam2022, AGOSTINELLI2022, LenoADRMP2020}.

\citet{Bosco2019} present a method to analyze UI logs in order to discover sequences of actions that are well defined and hence can be automated using RPA tools. The proposed method takes as an input a UI log that consists of a set of UI event sequences, i.e., routines. Each routine trace consists of a sequence of interactions (actions). Each action has a type (copy, paste, select, etc.) and a set of parameters. Given a UI log, the method outputs routine details and information.

\begin{enumerate}
\item
  Each candidate automatable routine is analyzed by checking whether
  each of its actions is deterministic (i.e., the action could be
  executed in a systematic way by RPA scripts) or not. The output is a
  tuple consisting of an action and a set of functions to automatically
  determine all the action's parameter values.
\item
  The maximal sequences of deterministic actions are extracted from the
  candidate automatable routines, and for each of them the activation
  condition of the first action is discovered.
\item
  The final output is a set of routine specifications. Each routine
  specification is a tuple consisting of an activation condition to
  automate the routine, and a sequence of action specifications.
\end{enumerate}

\citet{LenoADRMP2020} developed a robotic process mining (RPM) family of techniques that help determine which routines should be automated before starting an RPA deployment. RPM aims to discover the repetitive routines in a process that are suitable for automation from user interaction event logs. Additionally, the RPM tool includes several preprocessing steps that help discover the RPA scripts suitable for automation generated from event logs. This work is more focused on the preprocessing of the event logs collected after the recordings are done.
In terms of the project lifecycle, this paper covers steps S4 to S6. The first step in an RPM pipeline is to record the interactions between one or more workers and one or more software applications. The recorded data is represented as a sequence of user interactions such as copying a cell, pasting the copied data into a form, editing a text field in a form, etc. The unnecessary steps are filtered from the event logs. The second step is to decompose the event logs into segments. Those discovered segments are used to identify routines that are then analyzed to identify those that can be automated and then to encode them as RPA scripts.

The RPM pipeline is followed by \citet{Leno2020IdentifyingCR} to identify the candidate routines that can be automated from unsegmented event logs collected from user interaction activities. The approach used in this paper is composed of two macro steps: 1)~decompose the normalized UI log into segments, and 2)~identify candidate routines by mining sequential patterns from those segments. The paper proposes a method to split an unsegmented UI log into a set of segments, each representing a sequence of steps that are repeated in the unsegmented UI log. The first step is to segment the event log and generate control-flow graphs derived from the log. Then, pattern mining is used to discover frequent candidate routines for automation. The patterns are then ranked according to four quality criteria: frequency, length, coverage, and cohesion. This approach was extended by \citet{LENO2022101916} to present an approach to post-process the identified candidate routines in order to assess whether these routines can be automated or not. If the routine is fully automatable, then an executable routine specification can be generated. Additionally, the authors proposed a method to identify equivalent routines, which enables producing a non-redundant set of automated routines.

\citet{RPM2022} built on previous work and introduced Robotic Process Mining, a family of techniques to discover repetitive routines that can be automated using RPA. This approach consists of three phases:
\begin{enumerate}
\item
  Collecting and preprocessing UI logs.
\item
  Discovering candidate routine for RPA implementations.
\item
  Discovering executable RPA routines.
\end{enumerate}
The goal of Robotic Process Mining is to identify the sequences of UI that are repetitive based on multiple traces collected, and then identify the routines that are good candidates for automation.

\citet{AGOSTINELLI2022} explored several issues with the segmentation of UI event logs. The focus was on segmenting three different forms of UI logs: 1)~same routine with different executions, 2)~multiple executions for several routines without having common user actions, and 3)~multiple executions for several routines with the possibility of common user actions. These authors studied how the segmentation techniques behave in each these three cases. Their focus is again on preprocessing and restructuring the event logs to discover the routines that can be automated. Even though this paper did not focus on using process mining algorithms to discover the different paths of the processes, they still studied how to simplify the UI logs to discover segmentations that can help implement RPA.

\citet{choi2021} provided an approach for selecting candidate tasks for robotic process automation based on user interface logs and process mining techniques. Their approach also considers collecting and generating user interface logs. However, their approach focuses on the transformation stage for the UI logs, which is then followed by log filtering, and finally tasks discovery using PM. The authors implemented an approach consisting of transformation rules that are defined based on the information available in the user event logs. These rules are used to transform the name of the original action (for example ``Open'') into an action name including sufficient information (for example ``Open System Folder Orders'') for discovering tasks model describing the user's sequence of actions that can be automated. The transformation rules differ based on the action type. For example, the action ``Open'' will require a source type and a source name to know exactly what is opened, whereas the action ``Click Button'' needs information on the context to know what has been clicked.

\citet{Neelam2022} recently looked into the process selection strategies for RPA projects. Their main focus was on understanding which processes can be automated and what qualities are ideal to detect such processes, in addition to understanding the order in which the processes should be deployed using RPA. To answer their research questions, these authors focused on the three main techniques, namely RPA, a process quantifiable-based approach, and survey-based strategies.

Several other research contributions have discussed pattern mining and other techniques to simplify and restructure event logs. For example, \citet{ElGharib-API-2022} proposed a tool-supported methodology and an API for preprocessing event logs in order to simplify them by replacing patterns of events. This methodology can be applied in the context of RPA to discover the most frequent paths in a process. Several preprocessing algorithms have been explored in the domain of process mining research to simplify, preprocess, and discover paths from complex event logs, or to handle aspects such as privacy and anonymity (see \citet{RAFIEI2021101908}), but they have not been framed in the RPA context (e.g., \citet{eventlogprep2021}).

\subsubsection{Process discovery and analysis}
\label{subsect:discovery-analysis}
After the preprocessing and restructuring of the user interaction logs, process maps are generated using process mining tools to enable analysis. Step S6 in Figure~\ref{fig:steps} represents process discovery using PM algorithms. Sixteen papers used open-source and commercial tools to discover the process maps that represent the user interaction event logs. Section~\ref{sect:tools} summarizes the tools that are used in these papers.

\citet{Pareto-DATA2020} used the notion of variability in process mining to select the processes for automation. The author focused on the pareto principle~\cite{Sanders1987THEPP} or the 80/20 rule, which means that 80\% of the outcomes usually come from 20\% of the causes. If the event log has a pareto-like distribution, then the regular or frequent paths can be identified for automation while the infrequent patterns can be filtered out. Based on the pareto principle, the process variant frequencies can be classified into three groups~\cite{Pareto-DATA2020}:

\begin{enumerate}
\item
  Regular, highly frequent subprocesses that should be automated in a traditional way in the information system.
\item
  Frequent, standardized subprocesses that can be automated by robots.
\item
  Infrequent and exceptional process behaviors that are still handled by humans.
\end{enumerate}

According to \citet{Pareto-DATA2020}, RPA aims to automate the second group of subprocesses that are rather frequent, repetitive, and simple, and where it is not cost-effective to change existing information systems.

\citet{Choi2022} used the term ``Tasks Discovery'' to build process models and discover the user actions that can be automated using RPA from transformed UI logs. The authors defined three criteria to select the candidate tasks that are suitable for automation:

\begin{enumerate}
\item
  Frequency: Creating a model that shows the frequency of each task, by applying case frequency techniques, and the frequency of transition from one task to another. Based on the result, the most frequent routines are selected for RPA.
\item
  Periodicity: The aim here is to identify and select frequent periodic cases, for example that are performed every Monday.
\item
  Duration: The goal in calculating the duration is to identify the cases and tasks that are taking a long time to be executed by employees and that can be automated by bots in a much faster time, at times in a fraction of a second.
\end{enumerate}

\citet{FabianRPAFlowcharts2022} took the approach of using process discovery techniques to create RPA flowcharts that are needed to program the bots based on UI logs. The prototype that was implemented included three main components that aim to generate RPA flowcharts. The first component (\textit{Adoption of UI logs}) focuses on getting the UI logs and preprocessing as in other approaches in previously discussed papers since UI logs are the starting point for discovering candidate routines for automation. The second component (\textit{Process discovery}) focuses on discovering the process models and process steps based on UI logs as inputs. The discovery algorithms use process mining discovery techniques. In the last component (\textit{Generation of RPA flowchart}), the output of the process discovery algorithms, which are the paths selected for automation, are translated into RPA flowcharts. Then, the generated RPA flowcharts are used to execute RPA implementations. The prototype was tested using UI logs from Microsoft Excel, Microsoft Edge, and the operating system. This paper does not discuss the selection criteria used for selecting the paths. Step S7, which is the process analysis, aims to assess and analyze the process maps generated by step S6.

\citet{PMandRPAP2P} applied process mining techniques on a procure-to-pay dataset to discover and analyze the process in order to utilize RPA implementation to improve the performance. The authors also applied process mining techniques to measure the efficiency of RPA for saving costs. In particularly they applied the fuzzy minor algorithms to discovery repetitive tasks with high frequency that are best suited for RPA and then measure the results of the implementations.

\citet{Kirchmer2019} focused as well on several criteria that should be checked before deploying RPA solutions, which include identifying the high impact business processes and verifying that the RPA technology fits to the solution. This last part is concerned with whether the processes are repetitive transactions, high-transactional volume processes, and stable and well-defined processes.

\citet{Manufacture2022} illustrated how Intelligent Process Automation (IPA) can be applied in the manufacturing industry. The framework that they proposed consists of five components. 
\begin{enumerate}
\item
  Create an IPA Implementation Roadmap.
\item
  From Process Discovery to Process Mining.
\item
  IA Analysis and RPA Estimation.
\item
  Define Suitable Solution Architecture.
\item
  RPA Design First Then AI Capability Integration.
\end{enumerate}

As part of the second component, PM is the automated applied to discover the process models based on extracted data. The authors emphasized the importance of applying process mining to automatically build as-is process model and improve the efficiency and speed of automation. Applying process mining to discover processes is a critical step as part of IPA because it is the key to understand current state processes based on data, streamlines the identification of processes that are suitable for automation and IPA can be costly if implemented without fully understanding the existing processes.

\citet{rinderle2021} investigated how process mining impact automation strategies and vice versa, also  with a focus on manufacturing processes. If there is an existing dataset, then process mining can support automation. On the other hand, automating manufacturing processes with a proper logging mechanism can yield to a good data collection that will increase the quality of process mining and analysis.

\subsubsection{RPA Configuration and Sustainability}
\label{subsect:RPA-config}
Successful process automation requires knowledge about the potential for automation, effective training of the software bots, and continuous monitoring for their performance, which is represented from steps S8 to S11 in Figure~\ref{fig:steps}.

Process mining is presented as a way to identify what can be automated using RPA. However, process mining should not only be used in the implementation phase. By continuously monitoring and observing human problem resolving capabilities, for example system errors or unexpected system behavior, RPA tools can adapt and handle non-standard cases (see \citet{Aalst2018}). Moreover, process mining can also be used to continuously improve the work between systems, robots, and people. Steps S8 to S11 respectively target configuring the RPA scripts, implementing the software robots, monitoring the software robots using PM techniques, and sustaining the RPA implementation.

\citet{Geyer-Klingeberg2018124} also highlight best practices for a successful RPA implementation where process mining can accelerate and improve this approach, taking into consideration that not all processes are suitable for RPA:

\begin{enumerate}
\item
  Selecting the appropriate process use case that can be automated.
\item
  Standardizing the processes before automation.
\item
  Prioritizing the processes that can be automated.
\item
  Monitoring the results continuously.
\item
  Establishing a central unit for automation in the organization.
\end{enumerate}

\citet{Geyer-Klingeberg2018124} presented an approach that uses process mining to enable effective RPA within process transformation. They showed how the Celonis tool aims to support organizations throughout the whole lifecycle of RPA projects. Their approach consists of three steps: 1)~Assess RPA potential using process mining by identifying the processes that are scalable, repetitive, and standardized; 2)~Develop an RPA application by training the RPA robots with existing workflows and comparisons between humans and robots; and 3)~Sustain the RPA implementation through continuous monitoring.

\citet{Halaska2020} focused on process flow and logic criteria for implementing an RPA solution. In particular the recommendations made for RPA implementation are based on common patterns that are discovered using the process flow and the login criteria for a certain process. Their approach also focuses on discovering business processes using process mining techniques based on the event logs collected from information systems. This helps find common patterns in a process that are suitable for RPA. They focused on measuring the productivity and efficiency of tasks without RPA automation that are affected by the resource allocations and the specific type of human resources assigned for each task. They measured the time it takes to complete a task in the common patterns based on the quantity of resources allocated for such a task. They concluded that from productivity and efficiency perspectives, it is recommended to start implementing RPA solutions for tasks that take the most amount of time to be completed. The analysts should be considering automating first the common patterns that maximize productivity and efficiency.

\subsubsection{PM and RPA from Start to End}
\label{subsect:PM-RPA-Start2End}

None of the papers covered all the PM+RPA steps (defined in Figure~\ref{fig:steps}) from start to end. From Table~\ref{tab:papers-analysis}, we observe that a cluster of papers focus on steps S1 to S5, including the UI data preprocessing to identify segments and candidate routines for automation, whereas a smaller cluster focused on steps S6 to S11. The majority of the papers cover the generation and preprocessing of event logs to select the routines that can be automated. The frequency of papers drops as we look at the papers that use process mining techniques \textit{after} the RPA implementations. 

Process mining techniques were used before starting the implementation of RPA in order to understand the steps of the process that can be automated using robots. Then, after the implementation and deployment of the robots, process mining was used to monitor the robots and evaluate their behavior.

In terms of the approaches that cover the most steps from start to end, we observe from Table~\ref{tab:papers-analysis} that 
\citet{Agostinelli2020} display the largest coverage by addressing S1 to S9, whereas \citet{LENO2022101916} cover S1 to S8 and \citet{FabianRPAFlowcharts2022} cover S2 to S9. Four contributions also cover steps S1 to S7~\cite{Jimenez2019,choi2021,humaninLoop2022,Cabello2022}.

\subsection{Tools}
\label{sect:tools}
The majority of the papers focused on collecting more event logs at the user interface level, which required using APIs to record the actions that were happening in Excel or Web browsers~\cite{LenoV2019AlEp,Jimenez2019,luka2022, AgostinelliInteractiveUI2021, Cabello2022, humaninLoop2022}.

Two papers from \citet{Agostinelli2020, AgostinelliInteractiveUI2021} implemented the SmartRPA tool using Python. This tool exploits event logs to automatically generate executable RPA scripts that can be automated using software robots. Another open-source Java command line tool was implemented to identify segments and then detect candidates that can be automated using RPA \cite{LenoADRMP2020}. One paper from \citet{AGOSTINELLI2022} used the SmartRPA tool that was implemented during their previous work. The table below summarizes

The ProM framework, developed by \citet{PROM2005}, is an open-source PM environment extensible through plugins. This framework is flexible in terms of input and output formats, and it allows the implementation of custom PM algorithms.

A ProM plug-in was used to generate process maps using collected UI logs as an input~\cite{Jimenez2019}. \citet{Cabello2020} also used ProM as their process mining tool to discover the as-is processes and the paths that can be automated, while the UiPath RPA tool~\cite{UiPath} was used to program the software robots that automate the tasks.

Five of the selected papers used Disco~\cite{Disco} to discover the frequent and most common routines in event logs that can be implemented using RPA solutions~\cite{Agostinelli2019ResearchCF,Agostinelli2020,Halaska2020,choi2021,PMandRPAP2P}). These papers imported the event logs to Disco to generate process maps and visualize as-is processes and the sequences in which the tasks are executed.

In three papers by \citet{LenoV2019AlEp, Bosco2019}, and \citet{Halaska2020}, the authors used Apromore~\cite{Apromore} to generate process maps from event logs that helped discovering patterns in the most frequently executed routines.

Only one paper used Celonis~\cite{Celonis} to discover process maps from event logs, which were analyzed by \citet{Geyer-Klingeberg2018124} to improve the process in place based on which sub-processes can be automated using RPA. Celonis was also used to monitor the processes after they were automated using RPA.

One paper from \citet{FabianRPAFlowcharts2022} used PM4PY~\cite{PM4PY}, a process mining library for Python, to implement the process discovery approach using the $\alpha$-miner algorithm. Another paper from \citet{Manufacture2022} used Python to implement the proposed IPA framework. In another paper, \citet{humaninLoop2022} also used Python to implement the proposed tool to help users identify segments from event logs.

In another paper, \citet{luka2022} used XES \cite{XES} to implement the data model for process-related UI logs.

Table~\ref{tab:tools-analysis} summarizes the tools that were used in each step of Figure~\ref{fig:steps}. The tools were primarily used to generate, preprocess, and segment the UI logs, as well as selecting and discovery routines that are suitable for RPA implementations.

\begin{table}[]
    \centering
    
    \begin{tabular}{l|l|l|l|l|l|l|l|l|l|l|l|l|l}
    \hline
    \textbf{Tool} & \rotl{S1: IS interaction} & \rotl{S2: recording} & \rotl{S3: generation} & \rotl{S4: preprocessing} & \rotl{S5a: segmentation} & \rotl{S5b: routines} & \rotl{S5c: select} & \rotl{S6: discovery} & \rotl{S7: analysis} & \rotl{S8: RPA config} & \rotl{S9: robot impl.} & \rotl{S10: monitoring} & \rotl{S11: sustain} \\
    \hline
    Apromore & \N & \N & \N & \N & \N &\Y &\Y &\Y & \N & \N & \N & \N & \N\\
    ProM & \N & \N & \N & \N & \N & \N & \N &\Y &\Y & \N & \N & \N &\N\\
    Disco & \N & \N & \N & \N & \N &\Y &\Y &\Y &\Y & \N & \N & \N &\N\\
    UiPath PM & \N & \N & \N & \N & \N & \N & \N &\Y &\Y & \N & \N & \N &\N\\
    Celonis & \N & \N & \N & \N & \N & \N & \N &\Y &\Y & \N & \N &\Y &\Y\\
    SmartRPA & \N & \N &\Y &\Y &\Y & \N & \N & \N & \N & \N & \N & \N &\N\\
    PM4PY & \N & \N &\Y &\Y &\Y & \N & \N & \N & \N & \N & \N & \N &\N\\
    XES & \N & \N & \N &\Y &\Y &\Y & \N & \N & \N & \N & \N & \N &\N\\
    \hline
    \end{tabular}
    
    \caption{Summary of tools usage for PM+RPA steps (\colorbox{green!25}{\checkmark} = Used, \colorbox{red!25}{\tiny\strut} = Unused).}
    \label{tab:tools-analysis}
\end{table}

Table~\ref{tab:tools} summarizes the tools that were used in the selected papers for this SLR. The majority of the tools are open source since this provides flexibility for the researchers to implement their algorithms. The existing PM and RPA tools in the market do not however provide the flexibility for the developers to preprocess data as needed. For example, existing PM tools and RPA solutions do not have the capability to apply segmentation on UI logs and select the candidates that can be automated.

\begin{table}[]
    \centering
    \small
    \begin{tabular}{p{0.3\textwidth}|p{0.62\textwidth}}
        \hline
        \textbf{Process Mining Tools} & \textbf{Papers} \\
        \hline
        Apromore~\cite{Apromore} & \citet{Bosco2019,LenoV2019AlEp,Halaska2020} \\
        ProM~\cite{ProM} & \citet{Jimenez2019,Cabello2020} \\
        Disco~\cite{Disco} & \citet{Halaska2020, Agostinelli2019ResearchCF, Agostinelli2020, choi2021, PMandRPAP2P}\\
        UiPath PM~\cite{UiPathPM} & \citet{Cabello2020} \\
        Celonis~\cite{Celonis} & \citet{Geyer-Klingeberg2018124} \\
        SmartRPA~\cite{SmartRPA} & \citet{Agostinelli2020,AgostinelliInteractiveUI2021} \\
        PM4PY~\cite{PM4PY} & \citet{FabianRPAFlowcharts2022} \\
        XES~\cite{XES} & \citet{luka2022} \\
        \hline
    \end{tabular}
    \caption{Tools used in the selected papers}
    \label{tab:tools}
\end{table}

\subsection{Challenges}
\label{sect:challenges}
There are many challenges in applying PM techniques on event logs collected at the user-interface level of an application in order to discover frequent patterns that can be automated using RPA.

\citet{Agostinelli2019ResearchCF} analyzed the RPA tools that are available on the market and then developed a classification framework to categorize the tools based on these dimensions: software architecture, coding features, recording facilities, self-learning, automation type, routine composition, and log quality. Based on their results, four research challenges were derived to inject intelligence into current RPA technology.

\begin{enumerate}
\item
  Intra-routine self-learning (segmentation): none of the RPA tools that   exist on the market can automatically understand which user actions   have to be considered inside the log, interpret the granularity, and identify which routines these logs belong to.
\item
  Inter-routine self-learning: the identification of candidate routines that can be automated with RPA is itself not automated. This step is done through interviews, direct observation of employees, and documentation reviews, which are time consuming and might lead to
  errors.
\item
  Automated generation of flowcharts from RPA logs: generating different flowcharts from RPA logs would facilitate the monitoring process for the robots after implementation and help testing whether the robots are executing the steps as they should.
\item
  Automated routines composition: current RPA solutions allow developing software robots for executing single and independent routines, not multiple dependent ones.
\end{enumerate}

The RPM pipeline of \citet{Leno2020RoboticPM} introduced to use process discovery techniques to identify candidate routines for RPA implementation, also helped identify several challenges. These challenges are classified at the PM level as they are related to the event logs collected before building RPA solutions. They include action recording, noise filtering, segmentation, simplification, routine extraction, executable routine discovery, and compilation.

\citet{Pareto-DATA2020} highlights another challenge, which is to match the user interaction log based on identifiers, usernames, keywords, and labels when the data is collected from multiple sources. For example, when collecting data from task mining techniques, recordings, and other monitoring tools, the challenge is to combine all these logs together and match the identifiers that refer to the same process performed by
the same user.

\citet{choi2021} focused on three challenges that need to be addressed in order to use process mining techniques properly to identify candidate tasks for automation using RPA: 1)~generating event log from recorded interactions with user interface, 2)~case identification (defining a case ID of a user interface log), and 3)~case duration calculation of tasks taking into consideration real-life situations.

\citet{rinderle2021} studied manufacturing processes and also observed several challenges, including 1)~connecting the machines to the process, 2)~involving humans with an active or a passive way, and 3)~collecting high quality data from manufacturing processes that are suitable for process mining.

\begin{figure}
    \centering
    \includegraphics[width=\columnwidth]{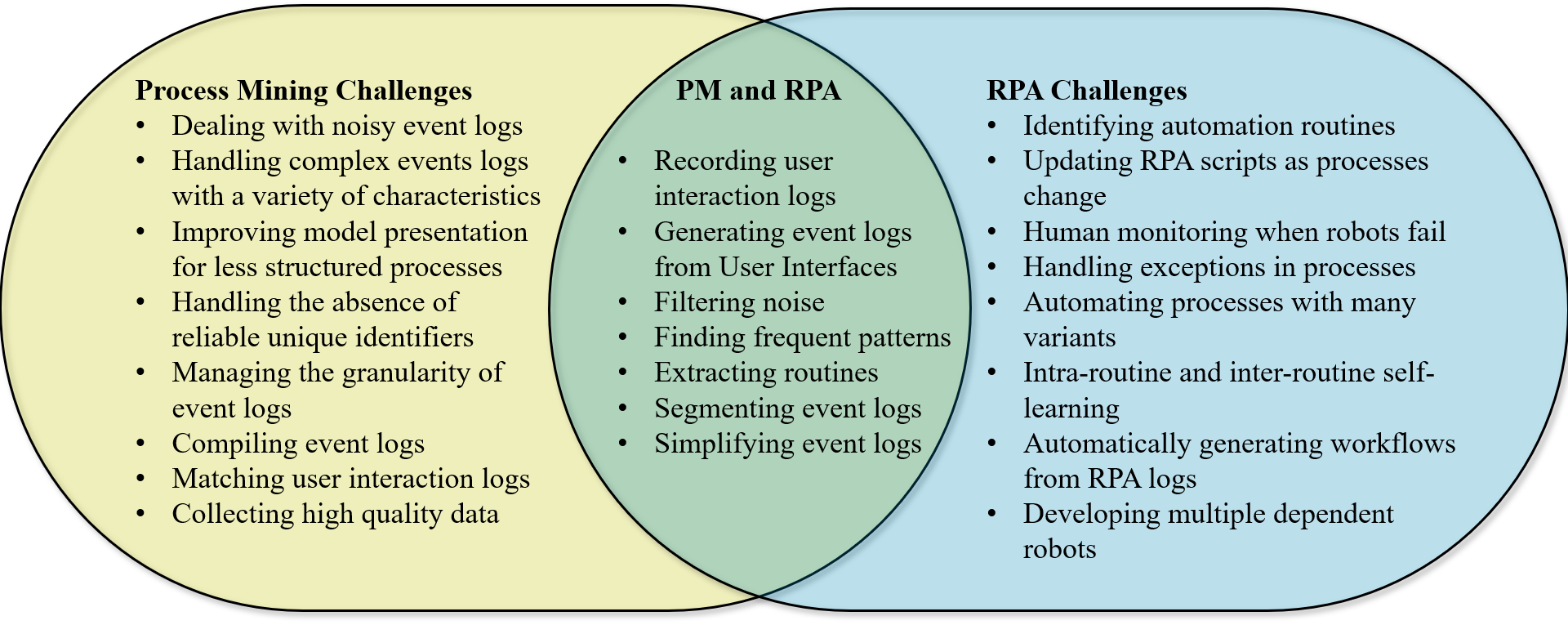}
    \caption{Process mining and RPA challenges}
    \label{fig:challenges}
\end{figure}

Figure~\ref{fig:challenges} highlights important process mining challenges in the context of RPA implementation, as well as general RPA challenges.

In addition to the process mining challenges that are discussed in the context of RPA, researchers have discussed other PM challenges that are not discussed in the RPA literature. Many of these challenges can also be applied to the RPA context. These challenges are related to the preprocessing and restructuring of event logs (steps S4-S5 in Figure~\ref{fig:steps}) to extract meaningful insights and generate process maps (step S6) that can be used in the analysis (step S7). \citet{ElGharibAmyotSLR} highlighted several challenges for applying PM in practice, which include:

\begin{enumerate}
\item
  Merging and cleaning raw event log data sets.
\item
  Dealing with complex event logs with multiple attributes.
\item
  Mining processes that change over time.
\item
  Improving the representation of process maps for less structured
  processes.
\item
  Mining complex processes with low levels of granularity (often
  collected from cloud-based applications).
\item
  Handling sequences of event where ordering does not matter.
\item
  Dealing with noisy event logs.
\end{enumerate}

These challenges have been discussed when applying process mining discovery algorithms in the context of cloud-based applications. Based on the analysis of the papers for this SLR, the challenges discussed by \citet{ElGharibAmyotSLR} are also relevant in the context of RPA when working with UI event logs in order to discover routines that are suitable for RPA implementation.

\section{Answers to the Research Questions}
\label{sect:answers}
To answer \textbf{RQ1} (Section~\ref{sect:RQs}) on how process mining techniques are applied to accelerate and improve robotic process automation, based on the 32 papers cited in this SLR, several results can be drawn. As there has been a major increase in RPA application across a wide range of industries that are automating processes, it is essential to know how processes are being executed and which ones can be automated. This is where process mining approaches show their strengths, as otherwise organizations often lack appropriate tools to understand their processes. 

Process mining can help analysts understand as-is processes and discover the routines that can be automated. To understand the processes that can be automated with RPA, it is essential to record the events that are executed by office staff to perform a task, including UI-related events. Then, recorded event logs give more details on the process insights when they are visualized with PM tools. Once RPA-based robots automating processes are deployed, PM can also provide continuous monitoring of these robots to ensure there are no errors or deviations. Further information is provided in Section~\ref{sect:techs-algos}, especially in terms of the steps that compose a typical PM+RPA project lifecycle (Figure~\ref{fig:steps}). In our SLR, the majority of the papers are using PM techniques to support the analysis and selection of processes that are suitable for automation with RPA. There is however a lack of papers that looked into how PM techniques can be used for handling the complexity coming with the use of UI logs, and for monitoring the performance of the robots and processes \textit{after} RPA implementations.

A combination of RPA and PM was used, for example, in the telecommunication industry at Vodafone \cite{Geyer-Klingeberg2018124}. PM raised alerts to Vodafone about many orders deviating from the expected or the standard process steps. These process variants with multiple deviations could not be automated with RPA. PM helped Vodafone identify those processes for improvement, which were then more easily automated using RPA without errors and with a faster speed. In this case, RPA helped Vodafone achieve a rate 92\% order rate leading to improvement on several metrics \cite{Geyer-Klingeberg2018124}.

\begin{tcolorbox}[left=2pt,right=2pt,top=0pt,bottom=0pt]
\textbf{Key message about RQ1:} Process mining techniques help discover routines that can be automated using RPA. Researchers should however tailor these techniques to support discovering processes based on UI logs, which is required for RPA implementations.
\end{tcolorbox}

To answer \textbf{RQ2}, several PM and RPA tools have been used in the selected papers (see Section~\ref{sect:tools}). The RPA tools that are in the market do not provide recording features to visualize the steps that are part of the process and to know how the processes happen in reality. Several PM tools, e.g., ProM, PM4PY, Disco, Apromore, and Celonis, have been used to build process maps using PM algorithms. These tools do not have the flexibility required to properly preprocess and simplify event logs in a way that would enable discovering the routines that can be automated. Assuming that proper preprocessing is done, PM tools can be used at the planning and assessment stage to discover the processes that can be automated and the sequence of events of a particular process. After the RPA software robots are programmed and are handling the process tasks, PM techniques can be used to monitor these automated processes to further discover errors, deadlocks, and deviations.

\begin{tcolorbox}[left=2pt,right=2pt,top=0pt,bottom=0pt]
\textbf{Key message about RQ2:} Based on the academic publications reviewed here, tool developers need to better integrate process mining and RPA features together, in a synergetic way. Functionalities of interest include preprocessing features to handle the complexity associated with UI logs, and features for monitoring end-to-end processes. Additionally, researchers should publish papers about opportunities and concrete benefits of such integrated tools.
\end{tcolorbox}

To answer \textbf{RQ3}, some of the selected papers in this SLR discussed the challenges encountered when combining process mining with robotic process automation. Before being able to apply PM discovery algorithms, the collected event logs require preprocessing. The list of challenges for PM, RPA, and their intersection are summarized in Figure~\ref{fig:challenges}, with more detail being provided in Section~\ref{sect:challenges}.
\begin{tcolorbox}[left=2pt,right=2pt,top=0pt,bottom=0pt]
\textbf{Key message about RQ3:} The research community should further investigate the challenges identified in Figure~\ref{fig:challenges}, especially those at the intersection of PM and RPA concerns, which tend to focus on data gathering and preprocessing.
\end{tcolorbox}

\section{Threats to Validity}
\label{sect:threats}
According to \citet{perry2000}, three types of threats to validity are relevant in literature reviews: 1)~construct validity, 2)~external validity, and 3)~internal validity.

\textbf{Construct validity} refers to the quality of the methodology in terms of being helpful to answer the research questions. Even though essential concepts and synonyms were included in the query that was ran across the most popular databases, in addition to using a snowballing approach, it could be possible that some relevant papers have not been found and were not included in this SLR. This is further exacerbated by the search being limited to peer-reviewed papers written in English. The grey literature was not reviewed and may include some additional information on how PM is being applied in practice in RPA projects. Such threats to validity were partially mitigated by doing a manual snowballing search to try including more papers that were not returned by the database searches.

\textbf{External validity} considers whether applying the conclusion of this study and the results to other cases or situations is possible. The focus of this SLR is limited to the use of process mining to accelerate and improve the implementation of robotic process automation. The methods, tools, results, and challenges that were discussed here were based specifically on the papers related to the intersection of PM with RPA. In particular, as papers exclusively focusing on the use of PM in support of RPA, the results may not generalize to other applications (e.g., on how RPA can support PM) or to other approaches (e.g., based on machine learning).

\textbf{Internal validity} examines any bias in performing the research. The major threat to the internal validity here is that the review of the literature was done mainly by the first author, with support from the second author. Some papers and analysis aspects thereof might have been overlooked or misrepresented due to fatigue or bias. This threat was partially mitigated by having a clear protocol first defined and validated by a peer, and a first version of this SLR peer-reviewed and validated by another peer.

\section{Conclusion}
\label{sect:conclusion}
The combination of process mining and RPA offers a unique opportunity to explore process management and to address the challenges of process discovery, improvement, and automation. Although RPA brings potential benefits to processes in terms of cost reduction and efficiency/effectiveness improvements, it is still important to carefully analyze the processes before considering their automation.
There is in particular no point in automating non-compliant or ineffective behavior~\cite{Pareto-DATA2020}. The application of PM is broader than RPA and does not stop after the software robots are deployed and operating. PM can be applied during the whole RPA project lifecycle starting with assessment (steps S2, S3, S4, S5a, S5b, and S5c in Figure~\ref{fig:steps}), software robots' implementation (steps S8 and S9), and process monitoring (steps S10 and S11).

This paper presented a systematic literature review on how process mining techniques and algorithms have been applied in the robotic process automation context to improve and accelerate automation implementation. Although there are reviews on PM and on RPA in isolation, none of them has focused on their intersection. 
RPA is an emerging technology that is designed to automate the repetitive, high-volume tasks that are executed by employees on their desktops by getting software robots to do them instead. It is essential to understand how these processes are executed in reality so they can be automated. PM techniques support the deployment of RPA projects by discovering as-is processes, which helps in selecting the processes and tasks that can be automated, with further support for their robot monitoring.

This SLR discussed and analyzed 32 peer-reviewed papers selected from database searches supplemented with snowballing. The contributions of this review include the identification of the techniques and algorithms that are currently used to improve and accelerate discovering and detecting the routines and the tasks that can be automated (RQ1). There is a focus on the techniques and algorithms to preprocess, prepare, restructure, and simplify the UI logs before importing such data sets into PM tools. A mix of open-source and commercial tools are being used at the PM and RPA levels (RQ2). This SLR also discusses the challenges that are encountered with combining PM with RPA during and automation project's lifecycle (RQ3). The answers to our research questions also include key messages to the researchers and tool developers interested in combining PM with RPA.

This review is of value to practitioners working either in the domain of process mining or robotic process automation, as it summarizes the knowledge and state-of-art techniques that lie at the intersection of both research domains. This review is also valuable to researchers and tool developers in the PM domain as it highlights further challenges that should be addresses at the event log level to better enable RPA solutions. Additionally, this review is important to the researchers in the RPA domain as it shows how PM can be leveraged in the RPA context and how it plays an increasingly significant role during all RPA stages. Both research areas have been growing recently, though there are still challenges that can be further explored in future research.

\section*{Acknowledgments}
The authors thank Ronqi Pan and Sepher Sharifi for their peer assessment of the SLR protocol and draft version, respectively, Amal Anda for useful comments, as well as the anonymous referees for insightful feedback and suggestions. This work was partially supported by NSERC (Discovery Grant) and the University of Ottawa (Admission Scholarship).



%%       \citet{<label>} ==> Jones et al. [21]
%%       \citep{<label>} ==> [21]
%%

\bibliographystyle{elsarticle-num-names} 


% \end{thebibliography}

\bibliography{reference}
\end{document}